\documentclass[
  aps,
  prfluids,
  onecolumn,
  amsmath,
  amssymb,
  floatfix
]{revtex4-2}

\usepackage{graphicx}
\usepackage{tabularx}
\usepackage{bbm}
\usepackage{bm}

\usepackage[hidelinks]{hyperref}

\usepackage[mathscr]{euscript} 
\usepackage{stackrel}

\usepackage{listings} \usepackage{courier} 
\usepackage{listings}
\usepackage{courier}   
\newcommand{\be}{\begin{equation}}
\newcommand{\ee}{\end{equation}}
\newcommand{\bea}{\begin{eqnarray}}
\newcommand{\eea}{\end{eqnarray}}

\begin{document}



\title{Dynamic Alignment or Angular Persistence?}

\author{Amir Jafari}
 \email{elenceq@jhu.edu} 
\noaffiliation


\begin{abstract}
Dynamic alignment in magnetohydrodynamic (MHD) turbulence is often inferred from a decrease of the amplitude-weighted mutual angle between Els\"asser increments toward smaller separations. We represent the logarithmic amplitude product and folded angle as a joint stochastic state and use finite-step transition probabilities to measure angular memory, signed angular drift, and mean-square angular displacement. The construction is tested in forced incompressible full MHD and balanced strong-guide-field reduced MHD. At fixed initial angle, large-amplitude increment pairs undergo smaller angular changes than small-amplitude pairs across the tested scale ranges, including states that begin at large angles. Normalized amplitude moments increase toward smaller separation, and stronger amplitude weighting gives smaller angular averages with stronger scale dependence. In the reduced-MHD data, large-angle, high-amplitude source populations contribute with the forward-transfer sign of the Politano--Pouquet third-order moment while undergoing comparatively small angular changes. The same amplitude ordering appears in physical time in the full-MHD data. A finite-time update based on the Els\"asser advective term closely tracks the conditional amplitude--angle dependence but does not isolate the scale-local cascade interaction. The Chapman--Kolmogorov discrepancy is smaller than the variation between spatial half-volumes but larger than sampling noise alone, indicating approximate rather than exact Markov behavior of the reduced state. The amplitude-weighted diagnostic therefore samples the joint amplitude--angle evolution rather than directly measuring population-wide rotation.
\end{abstract}

\maketitle
\section{Introduction}
\label{sec:introduction}

The standard Richardson--Kolmogorov phenomenology for strong magnetohydrodynamic (MHD) turbulence assumes that velocity and magnetic-field increments measured across separations $\ell_\perp$ perpendicular to the local magnetic field scale as $\ell_\perp^{1/3}$, corresponding to a H\"older exponent $h=1/3$ and the perpendicular energy spectrum $E(k_\perp)\sim k_\perp^{-5/3}$. Scale-dependent alignment was proposed as an explanation for a $k_\perp^{-3/2}$ spectrum \citep{Boldyrev2006}, and subsequent simulations reported spectra close to that form \citep{Mason2006,BoldyrevMasonCattaneo2009}. In incompressible MHD, the velocity $\boldsymbol{u}$ and magnetic field $\boldsymbol{B}$, expressed in Alfv\'en-speed units, define the Els\"asser fields $\boldsymbol{z}^{\pm}=\boldsymbol{u}\pm\boldsymbol{B}$, which are nonlinearly coupled through $(\boldsymbol{z}^{\mp}\cdot\nabla)\boldsymbol{z}^{\pm}$. The original alignment phenomenology concerns the angle between perpendicular velocity and magnetic fluctuations; here we use the folded angle between perpendicular Els\"asser increments, a related numerical diagnostic. The proposed angular reduction weakens the nonlinear interaction and modifies the constant-flux scaling.

Subsequent work placed alignment within an increasingly intermittent and amplitude-dependent description of MHD turbulence. In an intermittent cascade, a single H\"older exponent does not describe all fluctuations: different spatial sets may exhibit different local scaling exponents, so amplitude-conditioned populations need not evolve identically across scale. Conditional measurements and phenomenological models accordingly show that large-amplitude Els\"asser fluctuations can be more aligned and more anisotropic within the plane perpendicular to the local magnetic field, so different amplitude sectors need not contribute equally to conventional alignment statistics
\citep{ChandranSchekochihinMallet2015,
MalletSchekochihinChandran2015,MalletEtAl2016,
MalletSchekochihin2017}.
Related reconnection-mediated theories also give amplitude- and
statistical-order-dependent evolution of large-amplitude sheet-like
fluctuations \citep{MalletSchekochihinChandran2017}.
Numerical work has emphasized that different alignment diagnostics need
not have the same scale dependence \citep{Beresnyak2012}, while the
interpretation and asymptotic role of measured dynamic alignment remain
debated
\citep{Schekochihin2009,Beresnyak2011,Schekochihin2022}.
Recent solar-wind measurements likewise show strong dependence of
alignment statistics on fluctuation amplitude, gradients, and
Els\"asser imbalance
\citep{SioulasEtAl2024,SioulasEtAl2025}.
They motivate a direct comparison between the angular population and the
amplitude-weighted alignment diagnostic.

For a local mean-field direction $\widehat{\mathbf b}_r$, the
perpendicular projector and centered perpendicular Els\"asser
increments are defined by
\begin{align}
\mathbf P_r
&=
\mathbf I-
\widehat{\mathbf b}_r\widehat{\mathbf b}_r^{\,T},
\nonumber\\
\delta_{\mathbf r}\mathbf z^\pm_\perp
&=
\mathbf P_r
\left[
\mathbf z^\pm
\left(
\mathbf x+\frac{\mathbf r}{2}
\right)
-
\mathbf z^\pm
\left(
\mathbf x-\frac{\mathbf r}{2}
\right)
\right],
\qquad
\mathbf r\cdot\widehat{\mathbf b}_r=0 .
\label{eq:intro_projected_increments}
\end{align}
We will use the amplitude product and folded angular variables 
\begin{align}
A_r
&:=
\left|
\delta_{\mathbf r}\mathbf z^+_\perp
\right|
\left|
\delta_{\mathbf r}\mathbf z^-_\perp
\right|,
\nonumber\\
s_r
&:=
\sin\theta_r
=
\frac{
\left|
\delta_{\mathbf r}\mathbf z^+_\perp
\times
\delta_{\mathbf r}\mathbf z^-_\perp
\right|
}{
\left|
\delta_{\mathbf r}\mathbf z^+_\perp
\right|
\left|
\delta_{\mathbf r}\mathbf z^-_\perp
\right|
},
\qquad
0\leq\theta_r\leq\frac{\pi}{2}.
\label{eq:intro_state_variables}
\end{align}
Alignment and anti-alignment are thereby identified. Below,
$A_r$, $s_r$, and $\theta_r$ always refer to these locally
perpendicular increments. Their dependence on spatial position,
sampling direction, and stored field or time is left implicit unless
needed explicitly. The amplitude-weighted angular average of order $p$ is
\begin{equation}
\left\langle s_r\right\rangle_{A^p}
:=
\frac{
\left\langle A_r^p s_r\right\rangle
}{
\left\langle A_r^p\right\rangle
}
=
\left\langle s_r\right\rangle
+
\frac{
\operatorname{Cov}(A_r^p,s_r)
}{
\left\langle A_r^p\right\rangle
},
\qquad p\geq 0,
\label{eq:intro_weighted_decomposition}
\end{equation}
whenever the required moment exists. Thus
$\langle s_r\rangle_{A^0}=\langle s_r\rangle$, while for $p=1$
\[
\left\langle s_r\right\rangle_A
=
\frac{
\left\langle A_r s_r\right\rangle
}{
\left\langle A_r\right\rangle
}
\]
is the conventional sine-based dynamic-alignment measure.
Here and thereafter, unless explicitly marked by an amplitude
weight in the subscript, all averages and conditional averages are
unweighted. Scale dependence of $\langle s_r\rangle_A$ alone does not establish how
the unweighted angular population evolves. In the Richardson--Kolmogorov phenomenology, the nonlinear time is $\tau_{\rm nl}\sim r/\delta z_r$, so the constant-flux condition $\varepsilon\sim\delta z_r^2/\tau_{\rm nl}$ gives $\varepsilon\sim\delta z_r^3/r$, $\delta z_r\sim r^{1/3}$, and $E(k_\perp)\sim k_\perp^{-5/3}$. In the alignment phenomenology, angular weakening lengthens the nonlinear time to $\tau_{\rm nl}\sim r/(\delta z_r\sin\theta_r)$, and the same constant-flux condition becomes $\varepsilon\sim\delta z_r^3\sin\theta_r/r$. This relation alone cannot determine the scale dependences of both $\delta z_r$ and $\theta_r$, so an additional closure is required. In the original alignment argument, $\sin\theta_r\sim\delta z_r/v_A$, where $v_A$ is the guide-field Alfv\'en speed. Combining this closure with constant flux gives $\delta z_r\sim(\varepsilon v_A r)^{1/4}$, $\sin\theta_r\sim(\varepsilon r/v_A^3)^{1/4}$, and hence $E(k_\perp)\sim k_\perp^{-3/2}$ \citep{Boldyrev2006,Mason2006,BoldyrevMasonCattaneo2009}. The $-3/2$ spectrum follows from both the angular factor in the nonlinear time and the closure relating that angle to the fluctuation amplitude. The closure concerns the mutual angle of the interacting increments, whereas the measured diagnostic is amplitude weighted.

The original closure, however, is incompatible with the rescaling symmetry of reduced MHD. The guide-field speed enters the reduced-MHD equations only through the linear propagation operator $v_A\partial_\parallel$, so a change in $v_A$ can be absorbed into a rescaling of the parallel coordinate without changing the perpendicular dynamics. A universal perpendicular scaling may therefore involve $v_A$ only together with a parallel scale or time. The prediction $\delta z_r\sim(\varepsilon v_A r)^{1/4}$ instead depends on $v_A$ without such a parallel quantity and consequently does not respect this symmetry \citep{Beresnyak2012,Schekochihin2022}. This objection does not exclude scale-dependent alignment or an $r^{1/4}$ scaling; it shows that neither follows universally from the original closure. An RMHD-consistent aligned phenomenology must introduce additional information, such as a parallel outer scale through the combination $L_\parallel/v_A$, or an intermittent statistical description.

We describe the joint scale evolution of amplitude and angle through the
stochastic state
\begin{align}
\boldsymbol{\xi}_r
&:=
\left(
a_r,s_r
\right),
&
a_r
&=
\ln A_r,
&
\tau
&=
\ln\left(\frac{r_0}{r}\right),
\label{eq:intro_joint_state}
\end{align}
where $\tau$ increases toward smaller separation. The logarithm of
$A_r$ is used only as a convenient coordinate for its broad dynamic
range. For any pair of inertial-range separations $r_> > r_<$, the
two-scale joint probability
$p(\boldsymbol{\xi}_{<},\boldsymbol{\xi}_{>})$ contains the complete
statistical relation, within this reduced state space, between the
larger-scale state $\boldsymbol{\xi}_{>}=(a_>,s_>)$ and the
smaller-scale state $\boldsymbol{\xi}_{<}=(a_<,s_<)$. Conditioning this
joint probability on $\boldsymbol{\xi}_{>}$ gives the finite-step
probability of reaching each smaller-scale amplitude--angle state from
a prescribed larger-scale state. This construction requires neither a
continuous evolution in scale nor the identification of a scale
sequence with a material trajectory.

The conditional probability provides three angular statistics. Dependence of the destination distribution on $s_>$ measures angular memory. The signed conditional mean $\langle\Delta\theta\mid a_>,s_>\rangle$, with $\Delta\theta=\theta_<-\theta_>$, measures systematic angular motion; a negative value is a mean shift toward smaller folded angle. The conditional mean-square displacement $\langle(\Delta\theta)^2\mid a_>,s_>\rangle$ measures the total angular change. These statistics are evaluated across source amplitudes at fixed $s_>$. A decrease of the mean-square displacement with source amplitude defines amplitude-dependent angular persistence. The signed drift and the displacement magnitude are independent properties of the conditional probability, so a mean drift toward smaller angle can coexist with angular persistence.

The weighted measure at each separation follows from the corresponding one-scale distribution. Its finite change depends on the joint source--destination distribution because amplitude and angle changes must be paired within the same scale transition. Section~\ref{subsec:angular_persistence_formalism} gives the exact finite-step decomposition and the conditional angular moments used below.

We test the construction in the forced incompressible full-MHD simulation in the Johns Hopkins Turbulence Database (JHTDB)~\citep{JHTB1,JHTB2,Eyinketal2013} and in a balanced strong-guide-field reduced-MHD simulation~\cite{Beresnyak2014,Beresnyak2015}. The analysis includes conditional scale-space moments, one-scale amplitude statistics, Chapman--Kolmogorov composition and spatial transfer, a source-state Politano--Pouquet decomposition, and an Eulerian physical-time comparison with an Els\"asser advective-term update.

Section~\ref{sec:stochastic_formalism} develops the joint transition law and its angular statistics. Section~\ref{sec:numerical_tests} presents the scale-space, physical-time, transfer, and consistency tests. Section~\ref{sec:conclusions} summarizes the results and limitations. The appendices give the stochastic formulation and numerical details.

\section{Evolution of the amplitude--angle state}
\label{sec:stochastic_formalism}

The conventional diagnostic depends on both Els\"asser amplitude and mutual
angle. We describe their scale evolution through the joint state
$\boldsymbol{\xi}_r=(a_r,s_r)$, with $a_r=\ln A_r$ and
$s_r=\sin\theta_r$, as defined in Eq.~(\ref{eq:intro_joint_state}).
The scale coordinate
$\tau=\ln(r_0/r)$ increases toward smaller separations in the inertial range of turbulence.
Since $d\tau=-dr/r$, equal intervals in $\tau$ correspond to equal
fractional changes of $r$. 
  Consider two separations with $\tau_2>\tau_1$, so that $r_2<r_1$.
Without making any assumption about the evolution through scale, the
complete two-scale information contained in the reduced state is its
joint probability density
$p(\boldsymbol{\xi}_2,\tau_2;
\boldsymbol{\xi}_1,\tau_1)$. The one-scale distributions are the
corresponding marginals, obtained by integrating over the state at the
other scale. The same joint probability can be conditioned in either direction:
\begin{align}
p(\boldsymbol{\xi}_2,\tau_2;
   \boldsymbol{\xi}_1,\tau_1)
&=
K(\boldsymbol{\xi}_2,\tau_2
  \mid\boldsymbol{\xi}_1,\tau_1)
p(\boldsymbol{\xi}_1,\tau_1)
\nonumber\\
& =
K(\boldsymbol{\xi}_1,\tau_1
  \mid\boldsymbol{\xi}_2,\tau_2)
p(\boldsymbol{\xi}_2,\tau_2).
\label{eq:joint_kernel_factorization}
\end{align}
We use the
large-to-small-scale kernel because the physical question concerns the
evolution of the Els\"asser increments toward smaller
scales. This choice specifies the direction in which the cascade is
examined; it does not prevent conditioning on the same joint
probability in the opposite direction. The smaller-scale distribution is obtained from the larger-scale
distribution according to $
p(\boldsymbol{\xi}_2,\tau_2)
=
\int
K(\boldsymbol{\xi}_2,\tau_2
 \mid\boldsymbol{\xi}_1,\tau_1)
p(\boldsymbol{\xi}_1,\tau_1)
\,d\boldsymbol{\xi}_1$.
  The kernel maps the larger-scale state distribution to the smaller-scale distribution. 

Constructing a multiscale transition from neighbouring-scale kernels
requires a Markov condition.
For three ordered scales $\tau_1<\tau_2<\tau_3$, a Markov description
means that once the intermediate state $\boldsymbol{\xi}_2$ is known,
the larger-scale state $\boldsymbol{\xi}_1$ supplies no additional
information about the conditional distribution of
$\boldsymbol{\xi}_3$. The resulting composition law is
\begin{align}
&K(\boldsymbol{\xi}_3,\tau_3
   \mid\boldsymbol{\xi}_1,\tau_1)
\nonumber\\
&\quad =
\int
K(\boldsymbol{\xi}_3,\tau_3
  \mid\boldsymbol{\xi}_2,\tau_2)
K(\boldsymbol{\xi}_2,\tau_2
  \mid\boldsymbol{\xi}_1,\tau_1)
\,d\boldsymbol{\xi}_2 .
\label{eq:chapman_kolmogorov}
\end{align}
Equation~(\ref{eq:chapman_kolmogorov}) is the Chapman--Kolmogorov relation for a Markov process in scale: after conditioning on the intermediate state, the larger-scale state supplies no further information about the next step. Its formal properties are summarized in
Appendix~\ref{app:scale_stochastic}, while its numerical test is
described later with the numerical analysis.\footnote{This statistical Markov property should not be confused with locality
of the MHD nonlinear interaction. It concerns the information retained by the reduced state
$(a_r,s_r)$ and does not imply that the underlying nonlinear
interactions couple only neighbouring Fourier scales.}

If the Markov property holds over a sequence
$\tau_0<\tau_1<\cdots<\tau_N$, repeated composition gives
\begin{align}
&K(\boldsymbol{\xi}_N,\tau_N
   \mid\boldsymbol{\xi}_0,\tau_0)
\nonumber\\
&=
\int
\left[
\prod_{j=1}^{N-1}d\boldsymbol{\xi}_j
\right]
\prod_{j=0}^{N-1}
K(\boldsymbol{\xi}_{j+1},\tau_{j+1}
  \mid\boldsymbol{\xi}_j,\tau_j).
\label{eq:multiscale_kernel_path}
\end{align}
Equation~(\ref{eq:multiscale_kernel_path}) gives the probability of a
state at a more distant smaller scale, conditioned on the state at the
initial scale, after summing over all possible intermediate
amplitude--angle states. The kernels need not be identical at different
scales: the scale evolution can be inhomogeneous, so the ordered
sequence of transition probabilities must in general be retained. The sequence
$\boldsymbol{\xi}_0,\boldsymbol{\xi}_1,\ldots,\boldsymbol{\xi}_N$
is a path through scale. It is not the material trajectory of an
individual eddy or coherent structure; it describes the statistical
relation among matched Els\"asser increments as their measurement
scale is changed. 

A formal continuous limit follows by refining the scale partition. If
the interval between two fixed physical scales is divided into
progressively finer intervals in $\tau$, with the endpoint scales held
fixed, Eq.~(\ref{eq:multiscale_kernel_path}) formally becomes
\begin{align}
K(\boldsymbol{\xi}_f,\tau_f
  \mid\boldsymbol{\xi}_i,\tau_i)
&=
\int_{\boldsymbol{\xi}_i}^{\boldsymbol{\xi}_f}
{\cal D}\boldsymbol{\xi}\,
{\cal W}[\boldsymbol{\xi}] .
\label{eq:scale_path_integral}
\end{align}
Here ${\cal W}[\boldsymbol{\xi}]$ is the probability weight of a
continuous amplitude--angle history through scale. This representation
does not require a Langevin or Fokker--Planck model. If a continuous
scale generator exists, it determines ${\cal W}$; drift--diffusion is
only one special form of that generator. The general construction, its
functional representation, and its relation to the Kramers--Moyal
hierarchy and Pawula's theorem are given in
Appendix~\ref{app:scale_stochastic}.

\subsection{Angular persistence and amplitude-weighted alignment}
\label{subsec:angular_persistence_formalism}

The joint transition law separates angular persistence from systematic
rotation toward alignment. Let $r_>$ and $r_<$ denote any two
inertial-range separations with $r_> > r_<$, measured in the plane
perpendicular to the local magnetic field; $r_>$ is the source separation
and $r_<$ the destination separation of a generic scale transition.
Throughout, ``at fixed initial angle'' means conditioning on the angular
state $s_>$ at the source separation and comparing fluctuations with
different source amplitudes $a_>$ within that same angular state. In the
numerical analysis this conditioning is implemented using finite ranges of
$s_>$. The angle is not held fixed during the transition: its value at the
destination scale, $s_<$, is allowed to change and is the quantity whose
evolution we measure. We compare the angular changes of fluctuations with the same initial
angular state and different initial amplitudes using
\begin{align}
\mu_s(a_>,s_>)
&=
\left\langle
s_<\mid a_>,s_>
\right\rangle
-
\left\langle s_<\right\rangle ,
\nonumber\\
M_s^{(2)}(a_>,s_>)
&=
\left\langle
(s_<-s_>)^2
\mid a_>,s_>
\right\rangle ,
\nonumber\\
M_\theta^{(2)}(a_>,s_>)
&=
\left\langle
(\theta_<-\theta_>)^2
\mid a_>,s_>
\right\rangle .
\label{eq:angular_persistence_measures}
\end{align}
The quantity $\mu_s$ compares the conditional destination mean with the ordinary destination mean and therefore measures angular memory. The quantities $M_s^{(2)}$ and $M_\theta^{(2)}$ measure angular displacement in $s=\sin\theta$ and in the folded angle itself. No Gaussian form is assumed, and the full transition kernel retains higher conditional moments.

The direct-angle measure can be separated exactly into a systematic
mean change and the spread about that change:
\begin{align}
&M_\theta^{(2)}(a_>,s_>)
=
\left\langle
\Delta\theta
\mid a_>,s_>
\right\rangle^2
+
\operatorname{Var}
\left(
\Delta\theta
\mid a_>,s_>
\right),
\nonumber\\
&\quad
\Delta\theta=\theta_<-\theta_>.
\label{eq:angular_shift_spread}
\end{align}
where \[ \operatorname{Var}(\Delta\theta\mid a_>,s_>) = \left\langle \left[ \Delta\theta-\langle\Delta\theta\mid a_>,s_>\rangle \right]^2 \Bigm| a_>,s_> \right\rangle \] is the variance of the angular changes among fluctuations that begin in the same source state $(a_>,s_>)$. The first term in Eq.~(\ref{eq:angular_shift_spread}) describes systematic angular change, whereas the second
describes the spread of the angular changes around their conditional mean.
A mean shift toward smaller folded angle measures systematic alignment, whereas angular persistence concerns the displacement magnitude. The two can coexist: a population may drift toward smaller angle while its large-amplitude members undergo smaller total angular changes at fixed initial angle.

We use the term angular persistence when, at fixed initial angle, the
angular displacement decreases as the initial amplitude increases.
If large-amplitude fluctuations that begin at large angles
also undergo smaller angular changes, then large amplitude preserves
large-angle states as well as small-angle states. The amplitude
dependence cannot then represent a universal drift toward $\theta=0$.

The one-scale joint distribution also defines the normalized amplitude moments
\begin{equation}
I_p(r)
:=
\frac{\left\langle A_r^p\right\rangle}
{\left\langle A_r\right\rangle^p},
\qquad p>1 .
\label{eq:normalized_amplitude_moments}
\end{equation}
For a simply self-similar amplitude
$A_r=c(r)X$, with the normalized distribution of $X$ independent of
$r$, $I_p(r)$ is independent of scale. Growth of $I_p(r)$ toward smaller separation indicates increasing concentration of the $p$th moment in the large-amplitude tail. The weighted angular
average $\langle s_r\rangle_{A^p}$ defined in
Eq.~(\ref{eq:intro_weighted_decomposition}) uses the same factor
$A_r^p$. Increasing $p$ gives progressively greater weight to the angular states occupied by large-amplitude fluctuations. The pair $I_p(r)$ and $\langle s_r\rangle_{A^p}$ measures the scale-dependent association between amplitude intermittency and angular organization.

The two-scale joint probability also determines the change of the
conventional amplitude-weighted alignment measure
$\langle s_r\rangle_A
=\langle A_rs_r\rangle/\langle A_r\rangle$.
The higher-order quantities $\langle s_r\rangle_{A^p}$ introduced
above are used here as one-scale probes of intermittent
amplitude--angle organization; the finite-step attribution below is
applied specifically to the conventional $A$-weighted measure.
Define over one finite scale interval
$\Delta A=A_<-A_>$ and $\Delta s=s_<-s_>$. Its change can then be
written exactly as
\begin{align}
&\langle s\rangle_{A,<}
-
\langle s\rangle_{A,>}
=
\frac{1}{\langle A_<\rangle}
\bigg[
\left\langle A_>\Delta s\right\rangle
+
\left\langle
\bigl(
s_>-\langle s\rangle_{A,>}
\bigr)\Delta A
\right\rangle
+
\left\langle
\Delta A\,\Delta s
\right\rangle
\bigg].
\label{eq:weighted_finite_step_decomposition}
\end{align}

Equation~(\ref{eq:weighted_finite_step_decomposition}) is written using
the amplitude and angle at the larger separation as the starting values. The first term,
$\langle A_>\Delta s\rangle$, is the contribution from angular motion
itself: each fluctuation changes its angle by $\Delta s$, while being
weighted by the amplitude it already had at the larger separation. If the
amplitudes were otherwise unchanged, this would be the direct angular
contribution to the change of $\langle s\rangle_A$. The second term,
$\langle (s_>-\langle s\rangle_{A,>})\Delta A\rangle$, contains no angular
change. It measures how amplitude changes alter the relative statistical
weight of fluctuations that started at different angles. The third term,
$\langle\Delta A\,\Delta s\rangle$, is the contribution of amplitude and
angular changes occurring during the same scale transition. 

This form uses the larger-scale amplitude weights in the angular term. The three-term grouping is not unique: the cross
term can be combined algebraically with either of the first two terms
by using the smaller-scale amplitude or angle. Its magnitude refers to the particular decomposition in
Eq.~(\ref{eq:weighted_finite_step_decomposition}) and should not be
interpreted as a separate, uniquely defined physical mechanism. If a continuous Markov generator exists, the infinitesimal counterpart
of the cross term is described by the carr\'e du champ operator; this
analytic connection is given in
Appendix~\ref{app:carre_du_champ} \cite{BakryGentilLedoux2014}.

The three terms separate angular change, amplitude reweighting, and their joint change for this weighting convention. The weighted alignment measure is one projection of the joint transition law.

\subsection{Time evolution}
\label{subsec:temporal_formalism}

At fixed separation $r$, perpendicular to the mean magnetic field, the two-time joint
probability of $\boldsymbol{\xi}(t)$ and
$\boldsymbol{\xi}(t+\Delta t)$ determines the distribution of later
amplitudes and angles associated with a specified initial state. A
single finite-time comparison requires no assumption that the temporal
process is Markovian.

For source states with approximately the same mutual angle and different amplitudes, we compare the angular change over a finite interval $\Delta t$. If the
angular persistence found across scale reflects a property of
large-amplitude Els\"asser-increment pairs themselves, then, at the same
initial angle, large-amplitude pairs should also undergo smaller angular
changes in time than small-amplitude pairs. We test the same amplitude ordering in physical time without identifying time evolution with scale evolution. For incompressible MHD with equal viscosity and resistivity, the
Els\"asser equations may be written
\begin{align}
\partial_t\mathbf z^\pm
&=
-
(\mathbf z^\mp\cdot\boldsymbol{\nabla})\mathbf z^\pm
-
\boldsymbol{\nabla}P
+
\nu\nabla^2\mathbf z^\pm
+
\mathbf f^\pm .
\label{eq:elsasser_evolution_theory}
\end{align}
The local Els\"asser advective term contributes to both increment amplitude and mutual angle; pressure, dissipation, forcing, and changes of local geometry also contribute. The finite-time comparison measures the part associated with the advective term without treating it as the complete evolution.

\subsection{Predictions}

The conditional statistics define the following tests. Progressive alignment requires a negative signed drift,
\[
\langle\Delta\theta\mid a_>,s_>\rangle<0,
\qquad
\Delta\theta=\theta_<-\theta_>.
\]
Angular persistence requires $M_s^{(2)}$ and $M_\theta^{(2)}$ to decrease with source amplitude at fixed source angle; persistence at large initial angle excludes an interpretation confined to states already near alignment. The one-scale statistics test whether growth of $I_p(r)$ is accompanied by smaller and more scale-dependent $\langle s_r\rangle_{A^p}$ at larger $p$. The finite-step decomposition tests the separate angular, amplitude-reweighting, and joint contributions to the conventional weighted statistic. The transition kernels are tested by Chapman--Kolmogorov composition through an intermediate scale and by transfer to disjoint spatial and temporal samples. In physical time, we test the same amplitude ordering and compare the measured conditional changes with a finite-time update based on the Els\"asser advective term.

\section{Numerical Results}
\label{sec:numerical_tests}

We use two high-resolution numerical simulations with different magnetic-field
geometries. The full-MHD calculation uses the public
\texttt{mhd1024} simulation accessed through the Johns Hopkins
Turbulence Database \citep{JHTB1,JHTB2}. It is a forced, periodic,
pseudospectral incompressible MHD simulation on a $1024^3$ grid, with
$\nu=\eta=1.1\times10^{-4}$, $P_m=1$, Taylor--Green forcing at
$k_f=2$, $Re_{\lambda_u}\simeq186$, and no imposed uniform guide field.
The local magnetic geometry enters the measurement directly. At each separation the magnetic field is coarse-grained
locally, and the Els\"asser increments are evaluated in the
perpendicular plane defined by that field. Details of the filtering,
sampling volume, stored fields, interpolation, and numerical consistency tests are
given in Appendix~\ref{app:sampling_ck}. The scale-space measurements in \texttt{mhd1024} are matched at the
same spatial midpoint. A common azimuthal label identifies the sampling
direction across scales, while the local magnetic field, perpendicular
plane, physical separation direction, and centered increment endpoints
are reconstructed at each separation. The comparison uses matched equal-time measurements at different
separations.

The second numerical computation uses the balanced strong-guide-field RMHD
simulation described by
\citet{Beresnyak2014,Beresnyak2015}, specifically the $1024^3$
ordinary-viscosity run \texttt{b1024n}. The fluctuating velocity and
magnetic fields are transverse to the imposed guide field. Taking the
guide field along the $x$ direction, the increments are measured
directly in the fixed $y$--$z$ perpendicular plane, with the same
midpoint and perpendicular direction retained between the two
separations of each scale pair. The angular-persistence calculation uses the stored state
\texttt{R108000} as its primary state, with spatially disjoint samples
and \texttt{R162000} providing robustness tests. The
Politano--Pouquet calculation is repeated in five stored states,
\texttt{R018000}, \texttt{R072000}, \texttt{R108000},
\texttt{R126000}, and \texttt{R162000}. Further numerical details are
given in Appendix~\ref{app:beresnyak}.

We apply the angular-persistence diagnostics of
Section~\ref{sec:stochastic_formalism} to matched scale pairs in both
simulations. The one-scale intermittency statistics
$I_p(r)$ and $\langle s_r\rangle_{A^p}$ are evaluated separately over
nine separations using fifteen $320^3$ full-MHD subvolumes and the five
stored RMHD states R018000, R072000, R108000, R126000, and R162000.
For these ensemble statistics the uncertainties are estimated across
the fifteen full-MHD cubes or the five RMHD stored states, rather than
from the much larger number of spatial samples within each one. The
full-MHD scale sequence in the JHTDB also permits a direct test of the
finite-step transition description. Its densely sampled time sequence
is subsequently used to test angular persistence in physical time and
to compare the measured finite-time changes with an update based on the
Els\"asser advective term.
\begin{figure}
\centering
\includegraphics[width=0.95\textwidth]
{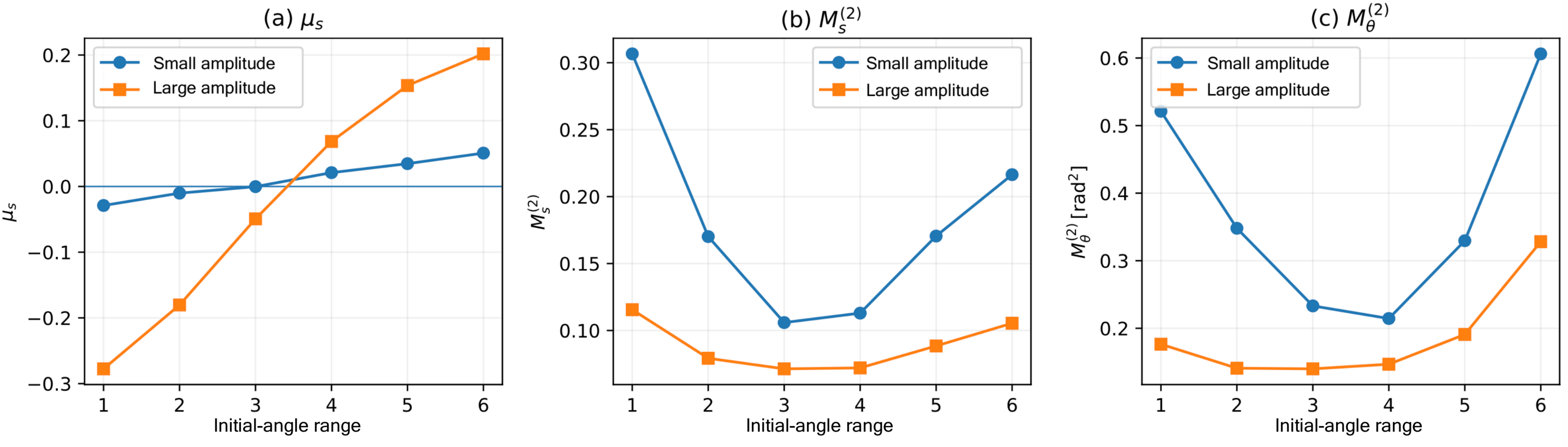}
\caption{
Scale-to-scale angular persistence of the Els\"asser increments in the
JHTDB for the representative transition
$r_{>}=63.5844\rightarrow r_{<}=55.4256$.
(a) Conditional angular memory $\mu_s$.
(b) Conditional mean-square change $M_s^{(2)}$.
(c) Corresponding folded-angle measure $M_\theta^{(2)}$.
At fixed initial angle, the large-amplitude population has smaller folded-angle changes. The same ordering is found throughout the tested scale range.
}
\label{fig:scale_space_summary}
\end{figure}

\subsection{Angular persistence across scale}
\label{subsec:scale_tests}

We first examine the scale-to-scale angular change of the perpendicular
Els\"asser increments in the JHTDB full-MHD simulation. Figure~\ref{fig:scale_space_summary} shows a representative transition
from $r_{>}=63.5844$ to $r_{<}=55.4256$. The conditional angular memory
in panel (a) retains the ordering of the larger-scale angular state:
measurements starting at small angle remain preferentially on the
small-angle side of the destination population, whereas those starting
at large angle remain preferentially on the large-angle side. The
memory is stronger at larger amplitude.

Panels (b) and (c) show smaller changes in both $s$ and the folded angle for the large-amplitude populations throughout the initial-angle distribution, including initially large-angle states.

The amplitude ordering is found in every initial-angle range at all
twenty-two tested starting separations.  In all 132
starting-scale--angle combinations, the rank dependence of both
$M_s^{(2)}$ and $M_\theta^{(2)}$ on source amplitude is negative, and
in every case the largest-amplitude range has smaller angular
displacement than the smallest-amplitude range.  A stricter
adjacent-range test gives the same result: $M_s^{(2)}$ decreases in
658 of 660 successive-amplitude comparisons and $M_\theta^{(2)}$ in
659 of 660.  The amplitude contrast is also substantial: the median ratios of the largest-amplitude to
smallest-amplitude angular displacement are 0.494 for $M_s^{(2)}$
and 0.497 for $M_\theta^{(2)}$.  Agreement between the two angular
measures shows that the result is not an artifact of using
$s=\sin\theta$ to represent the folded angle.

\begin{figure}
\centering
\includegraphics[width=0.95\textwidth]
{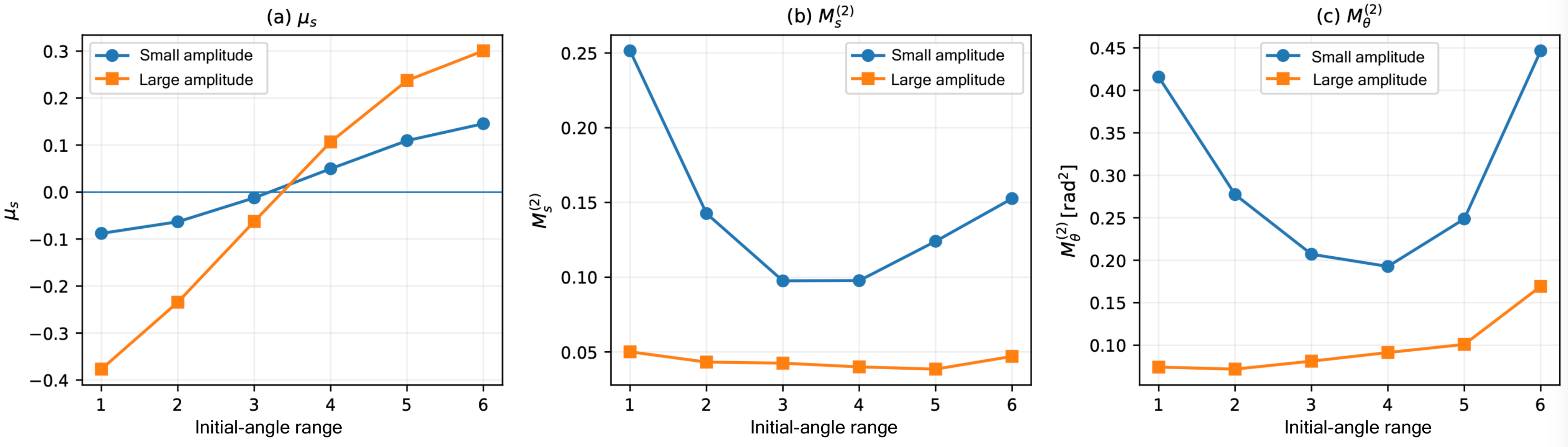}
\caption{
Scale-space angular persistence in the balanced strong-guide-field
RMHD simulation for the representative transition
$r_{>}=64\rightarrow r_{<}=56$, using the pooled perpendicular
directions.
(a) Conditional angular memory $\mu_s$.
(b) Conditional mean-square angular change $M_s^{(2)}$.
(c) Corresponding folded-angle measure $M_\theta^{(2)}$.
For both angular measures, the large-amplitude population undergoes
smaller angular displacement throughout the initial-angle distribution.
}
\label{fig:beresnyak_representative}
\end{figure}

The balanced RMHD simulation gives the same amplitude ordering. The scale
choices differ slightly between the two calculations because the JHTDB
analysis also requires a dense logarithmic sequence for the finite-step
composition test through intermediate separations. Its three-step persistence
transitions therefore satisfy
$r_{<}/r_{>}=3^{-1/8}\simeq0.872$, giving noninteger values such as
$r_{>}=63.5844$ and $r_{<}=55.4256$. The RMHD calculation is used as an
independent test of angular persistence and therefore uses individual
grid-aligned scale pairs. We choose
$r_{<}/r_{>}=7/8=0.875$, closely matching the full-MHD scale change while
keeping the centered increment endpoints exactly on grid points.
Figure~\ref{fig:beresnyak_representative} shows the representative transition
$r_{>}=64$ to $r_{<}=56$, with the two perpendicular directions pooled.
As in the full-MHD calculation, the large-amplitude fluctuations retain
stronger memory of their initial angular state and undergo smaller angular
displacement over the full initial-angle range.

The ordering holds over the full tested RMHD scale range. For the primary stored state,
both $M_s^{(2)}$ and $M_\theta^{(2)}$ decrease through every successive
amplitude interval at fixed initial angle in all 270/270 comparisons
for one perpendicular direction and independently in all 270/270
comparisons for the other. Spatially disjoint samples and the second
stored state recover the same large-amplitude versus small-amplitude
ordering; the detailed robustness tests are given in
Appendix~\ref{app:beresnyak}.
Both simulations give the same amplitude ordering despite their different perpendicular geometries. At fixed initial angular state, large-amplitude Els\"asser-increment pairs undergo less angular change across scale. Its extension to initially large-angle states identifies amplitude-dependent persistence of the angular configuration; a universal rotation toward alignment is a separate signed-drift question. A fixed vector change turns a larger increment through a smaller angle, so this ordering contains a geometric contribution.

We test Eq.~(\ref{eq:chapman_kolmogorov}), the Chapman--Kolmogorov composition law. If the reduced state $(a,s)$
contains the information needed for the finite-step scale evolution,
then the transition from an initial scale to a smaller scale should,
within finite-sample precision, agree with the transition obtained by
composing the two measured transitions through an intermediate scale. To estimate these transition probabilities from the data, we discretize
the joint amplitude--angle plane. For example, a $6\times6$ partition
means six equal-population ranges in $a=\ln A$ and six equal-population
ranges in $s=\sin\theta$, giving $36$ joint amplitude--angle states.
We repeat the calculation with $5\times5$, $6\times6$, and
$7\times7$ partitions, corresponding to $25$, $36$, and $49$ joint
states, to check that the result does not depend on one particular
discretization of $(a,s)$.

We quantify the difference between two transition kernels using the
source-weighted total-variation distance
\[
D_{\rm TV}(T,T')
=
\frac{1}{2}
\sum_{\beta}p_{\beta}
\sum_{\alpha}
\left|T_{\alpha\beta}-T'_{\alpha\beta}\right|,
\]
where $\beta$ labels the source amplitude--angle state, $\alpha$ the
destination state, and $p_{\beta}$ is the probability of the source
state. Thus $D_{\rm TV}=0$ means that the two conditional transition
distributions are identical, while larger values indicate increasing
disagreement. Define the Chapman--Kolmogorov discrepancy by
\[
D_{\rm direct-composed}
=
D_{\rm TV}(T_{\rm direct},T_{\rm composed}).
\] To determine
how large a discrepancy can arise simply from finite spatial sampling,
we divide the sampled volume into two disjoint spatial half-samples and
estimate the direct and composed kernels independently in each. We define
\begin{align}\nonumber
D_{\rm sampling}
&=
\frac{1}{2}
\Big[
D_{\rm TV}(T_{\rm direct}^{(1)},T_{\rm direct}^{(2)})\\\nonumber
&+
D_{\rm TV}(T_{\rm composed}^{(1)},T_{\rm composed}^{(2)})
\Big].
\end{align}
Thus $D_{\rm sampling}$ measures the variation between two disjoint
half-volume estimates. It is a conservative reference for spatial
variation, not the sampling noise of the full-sample
$D_{\rm direct-composed}$. Under exact Chapman--Kolmogorov composition,
a null calibration gives an expected ratio of about $0.45$ for the
full-sample discrepancy, or about $0.65$ when that discrepancy is averaged
over the two half-samples.
\begin{figure}
\centering
\includegraphics[width=0.852\textwidth]
{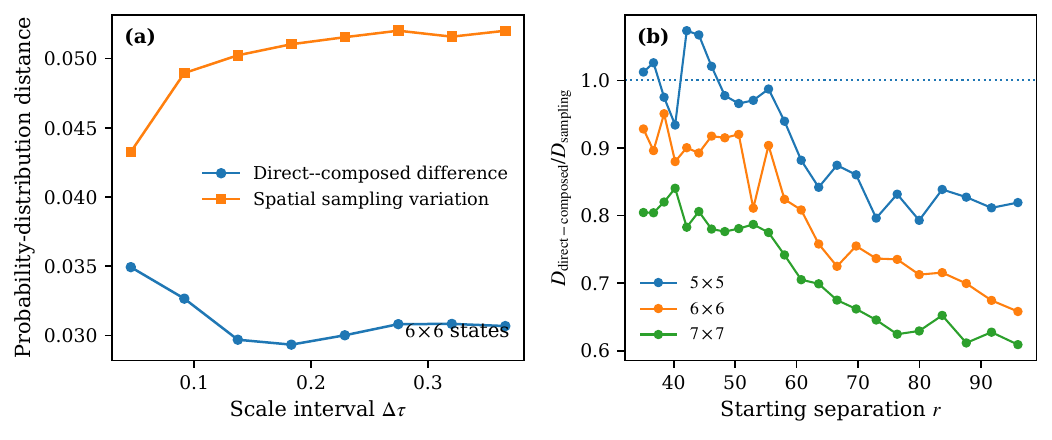}
\caption{
Chapman--Kolmogorov comparison for the finite-step scale evolution in
the JHTDB simulation. The joint state $(a,s)$ is divided into
equal-population amplitude and angle ranges; an $n\times n$ partition
therefore contains $n^2$ joint amplitude--angle states.
(a) Source-weighted total-variation distance
$D_{\rm direct-composed}$ between the directly measured transition and
the transition composed through an intermediate separation, compared
with the finite-spatial-sampling baseline $D_{\rm sampling}$, for the
$6\times6$ ($36$-state) partition.
(b) Ratio $D_{\rm direct-composed}/D_{\rm sampling}$ at the finest
tested scale interval for $5\times5$ ($25$ states), $6\times6$
($36$ states), and $7\times7$ ($49$ states) partitions.
The dotted line marks equality with the half-volume variation. Values
below unity do not by themselves establish consistency with sampling
noise.
}
\label{fig:finite_step_ck}
\end{figure}
Figure~\ref{fig:finite_step_ck}(a) compares these two quantities for the
$6\times6$ ($36$-state) partition as the scale interval is increased.
Here $\Delta\tau$ is the separation in logarithmic scale between
successive members of the three-scale sequence; the direct transition
therefore spans $2\Delta\tau$. At the finest tested interval, $\Delta\tau=0.0457755$, the mean
direct--composed differences are $0.02783$, $0.03494$, and $0.04201$
for the $25$-, $36$-, and $49$-state partitions, respectively. The
corresponding spatial-sampling baselines are $0.03047$, $0.04327$, and
$0.05849$. For all three discretizations, the mean direct--composed discrepancy is
smaller than the half-volume variation. The corresponding ratios are
$0.91$, $0.81$, and $0.72$ for the $25$-, $36$-, and $49$-state
partitions, respectively, all above the null expectation.

Figure~\ref{fig:finite_step_ck}(b) gives the more stringent
starting-separation test. For the $36$- and $49$-state partitions, the
direct--composed discrepancy remains below the spatial-sampling baseline
at every tested starting separation. The $25$-state partition has five
marginal exceptions toward the smaller-separation end of the range, with
a maximum ratio of $1.074$. The departure from Chapman--Kolmogorov
composition is smaller than the variation of the kernels between spatially
disjoint half-volumes, but larger than sampling noise alone would give; the
reduced state is approximately, not exactly, Markov at this resolution. The
detailed construction and additional scale intervals are given in
Appendix~\ref{app:sampling_ck}.

As a separate out-of-sample check, transition probabilities estimated
in one spatial region retain predictive information when applied,
without refitting, to a spatially disjoint region. The transfer test
is described in Appendix~\ref{app:sampling_ck}.

\begin{figure*}[t]
\centering
\includegraphics[width=0.97\textwidth]
{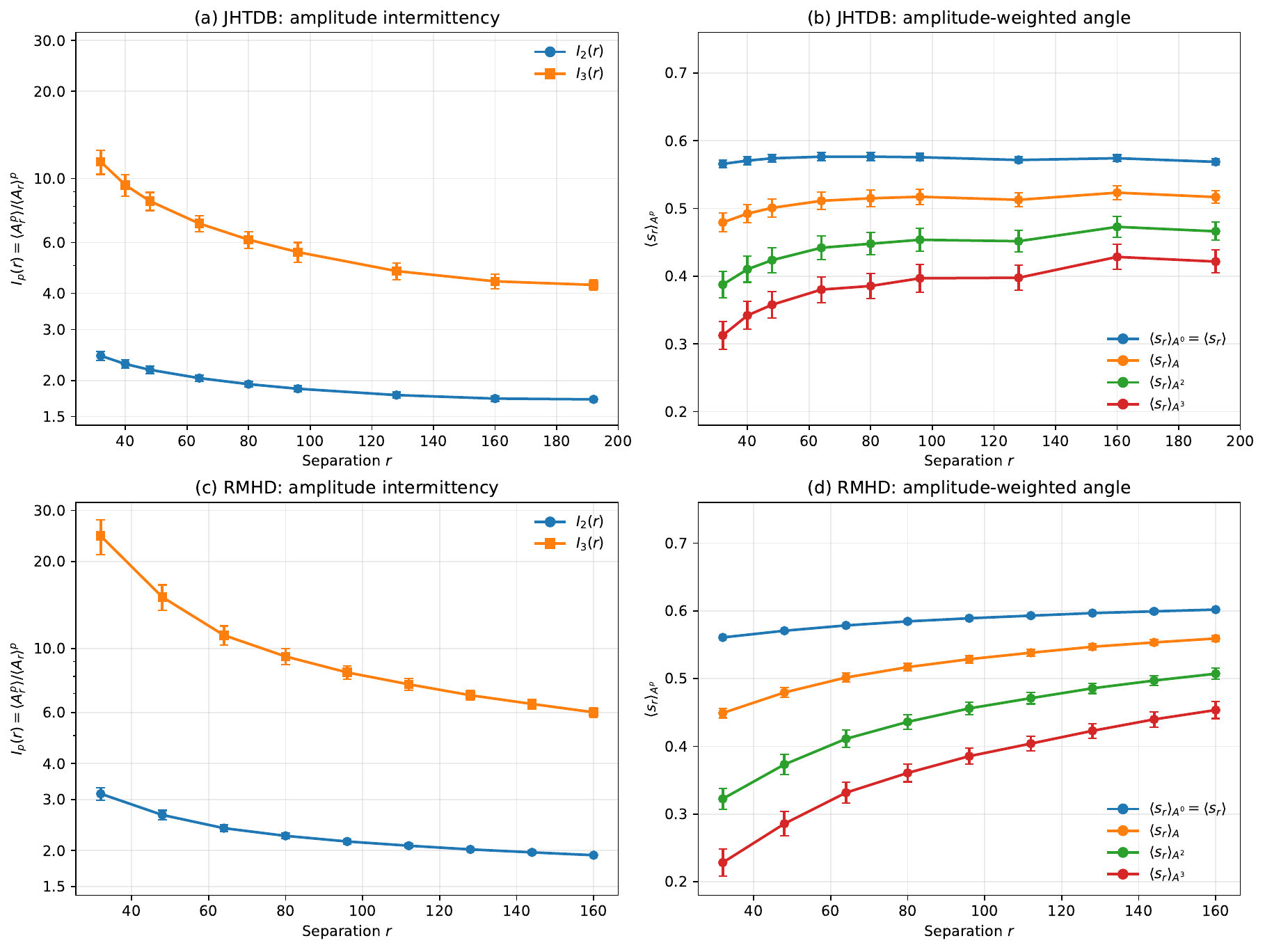}
\caption{\footnotesize
Amplitude intermittency and amplitude-weighted angular statistics in
the full-MHD JHTDB ensemble (top) and the balanced strong-guide-field
RMHD simulation (bottom). Panels (a) and (c) show the normalized
amplitude moments
$I_p(r)=\langle A_r^p\rangle/\langle A_r\rangle^p$
for $p=2$ and $3$. Panels (b) and (d) show the corresponding angular
hierarchy $\langle s_r\rangle_{A^p}$ for $p=0,1,2,3$, with
$\langle s_r\rangle_{A^0}=\langle s_r\rangle$ and
$\langle s_r\rangle_A$ the conventional dynamic-alignment statistic.
Toward smaller separation, the normalized amplitude moments increase
while the scale dependence of the angular statistic becomes
progressively stronger with increasing amplitude-weighting order.
Error bars denote standard errors across the fifteen full-MHD cubes
and the five stored RMHD states, respectively.
}
\label{fig:higher_moment_alignment}
\end{figure*}

The one-scale joint statistics show more directly how the conventional
weighted diagnostic is embedded in the intermittent amplitude--angle
organization. Figure~\ref{fig:higher_moment_alignment} shows the
normalized amplitude moments $I_p(r)$ together with the hierarchy
$\langle s_r\rangle_{A^p}$.

In the fifteen-cube full-MHD ensemble, $I_2$ increases from $1.720$ at
$r=192$ to $2.434$ at $r=32$, while $I_3$ increases from $4.281$ to
$11.421$. Over the same range the unweighted angular statistic is
nearly unchanged:
$\langle s_r\rangle_{A^0}$ changes from $0.569$ to $0.566$.
In contrast, $\langle s_r\rangle_A$ decreases from $0.517$ to $0.479$,
$\langle s_r\rangle_{A^2}$ from $0.466$ to $0.388$, and
$\langle s_r\rangle_{A^3}$ from $0.422$ to $0.313$.
Thus the scale dependence of the angular statistic becomes
progressively stronger as the amplitude-weighting order is increased.
Increasing $p$ progressively emphasizes the high-amplitude tail,
while the growth of $I_p(r)$ toward smaller separation shows that the
corresponding amplitude moments become increasingly concentrated in
that tail. The fifteen-cube values and their standard errors are
obtained by treating the cubes, rather than the individual spatial
samples, as the ensemble units.

The balanced RMHD calculation shows the same organization more
strongly. Between $r=160$ and $r=32$, $I_2$ increases from $1.927$ to
$3.143$ and $I_3$ from $6.015$ to $24.520$. Over the same interval,
$\langle s_r\rangle_{A^0}$ decreases from $0.602$ to $0.561$, while
$\langle s_r\rangle_A$, $\langle s_r\rangle_{A^2}$, and
$\langle s_r\rangle_{A^3}$ decrease respectively from $0.559$,
$0.507$, and $0.454$ to $0.449$, $0.322$, and $0.228$. In both
simulations the ensemble means therefore satisfy
$\langle s_r\rangle_{A^0}>
\langle s_r\rangle_A>
\langle s_r\rangle_{A^2}>
\langle s_r\rangle_{A^3}$
throughout the plotted scale range, with the separation between
weighting orders becoming substantially larger toward smaller scales.

The normalized amplitude moments and weighted angular means form a common hierarchy: larger $p$ emphasizes an increasingly intermittent large-amplitude sector concentrated at smaller $s$.

To determine what produces the decrease of the conventional
$A$-weighted statistic, we evaluated the three contributions in
Eq.~(\ref{eq:weighted_finite_step_decomposition}) for every measured
scale transition for which $\langle s\rangle_A$ decreases.  In the two spatially disjoint full-MHD samples and in the
RMHD calculation, the joint term
$\langle\Delta A\,\Delta s\rangle/\langle A_<\rangle$ is negative and
is the largest contribution in magnitude in all 50/50 such scale
transitions.  By contrast, neither the direct angular term
$\langle A_>\Delta s\rangle/\langle A_<\rangle$ nor the pure
amplitude-reweighting term
$\langle(s_>-\langle s\rangle_{A,>})\Delta A\rangle/
\langle A_<\rangle$
is negative in any of these transitions.  The median fractions of the
sum of the absolute magnitudes of the three terms are approximately
0.20, 0.23, and 0.56 for the direct, reweighting, and joint terms,
respectively, with nearly the same values in the three data sets.
Neither angular change at fixed amplitude weights nor amplitude
reweighting at fixed angle produces the decrease; it requires their joint
change. The same signs can arise without any scale trend when the joint
statistics at the two separations are interchangeable, so the decrease is a
small remainder of larger opposing terms.

\subsection{Politano--Pouquet relation and angular persistence}
\label{subsec:pp_persistence}

The Politano--Pouquet (PP) relation connects a third-order moment of the
Els\"asser increments to the mean rate at which Els\"asser energy is
transferred toward smaller scales. For homogeneous incompressible MHD
in a stationary inertial range \citep{PolitanoPouquet1998},
\begin{equation}
\boldsymbol{\nabla}_{\boldsymbol r}\boldsymbol{\cdot}
\left\langle
\delta\boldsymbol z^{\mp}
\left|\delta\boldsymbol z^{\pm}\right|^2
\right\rangle
=
-4\varepsilon^{\pm},
\label{eq:pp_divergence}
\end{equation}
where $\varepsilon^\pm$ is the mean transfer rate of the corresponding
Els\"asser energy. In RMHD, the fluctuating Els\"asser fields are
perpendicular to the guide field. If their statistics are isotropic
within the perpendicular plane, integration over that plane gives
\citep{Galtier2011}
\begin{equation}
\left\langle
\delta z_{L,\perp}^{\mp}
\left|\delta\boldsymbol z_{\perp}^{\pm}\right|^2
\right\rangle
=
-2\varepsilon^{\pm}r_{\perp},
\label{eq:pp_perpendicular}
\end{equation}
where
$\delta z_{L,\perp}^{\mp}
=\delta\boldsymbol z_{\perp}^{\mp}\boldsymbol{\cdot}
\widehat{\boldsymbol r}_{\perp}$.
The factor $2$, rather than the three-dimensional isotropic factor
$4/3$, results from integration in the two-dimensional perpendicular
plane. The negative sign corresponds to transfer from larger to smaller
perpendicular scales.

We test Eq.~(\ref{eq:pp_perpendicular}) without conditioning on
amplitude or angle. At each stored state and each separation
$r_\perp$, we calculate
$\delta z_{L,\perp}^{\mp}
|\delta\boldsymbol z_{\perp}^{\pm}|^2$
at all sampled spatial midpoints and average the result. For a
separation along $y$, the longitudinal increment is the $y$ component;
for a separation along $z$, it is the $z$ component. The two separation
directions and the two Els\"asser relations are examined separately.
Equation~(\ref{eq:pp_perpendicular}) predicts that the third-order
moment is negative and approximately proportional to $r_\perp$. Equivalently, Eq.~(\ref{eq:pp_perpendicular}) gives
\begin{equation}
-\frac{
\left\langle
\delta z_{L,\perp}^{\mp}
|\delta\boldsymbol z_{\perp}^{\pm}|^2
\right\rangle
}{2r_\perp}
=
\varepsilon^\pm .
\label{eq:pp_scale_estimate}
\end{equation}
We test whether the left-hand side of
Eq.~(\ref{eq:pp_scale_estimate}) is approximately independent of
$r_\perp$. The integration required
to obtain Eq.~(\ref{eq:pp_perpendicular}) has already been performed
analytically; the numerical calculation directly measures the
third-order moment at each separation. Because $\varepsilon^\pm$ is
not measured independently here, we test the predicted sign and scale
dependence but do not use Eq.~(\ref{eq:pp_scale_estimate}) as an
independent absolute measurement of $\varepsilon^\pm$.

The PP moment is a signed longitudinal quantity and cannot be replaced
by a positive expression involving only the increment amplitudes and
$\sin\theta$. The mutual angle $\theta$ is the angle between
$\delta\boldsymbol z^+$ and $\delta\boldsymbol z^-$, whereas
$\delta z_{L,\perp}^{\mp}$ is the projection of one increment onto the
separation direction. The PP relation does not determine
whether the mutual angle increases or decreases toward smaller
separation.

After testing the unconditioned relation, we divide the samples at a
larger source separation $r_>$ according to their source amplitude
$A_>$ and source angle $\theta_>$. The source angle is first divided
into equal-population ranges, and the amplitude is then divided into
equal-population ranges within each angular range. Let
$\mathcal C_{ij}$ denote the resulting source-state cell with angular
index $j$ and amplitude index $i$. For every cell, we calculate the
conditional mean of the same signed third-order quantity. The
unconditioned PP moment is then recovered exactly from
\begin{align}
\left\langle
\delta z_{L,\perp}^{\mp}
|\delta\boldsymbol z_{\perp}^{\pm}|^2
\right\rangle
={}&
\sum_{i,j}p_{ij}
\left\langle
\delta z_{L,\perp}^{\mp}
|\delta\boldsymbol z_{\perp}^{\pm}|^2
\middle|
\mathcal C_{ij}
\right\rangle ,
\label{eq:pp_source_state_decomposition}
\end{align}
where $p_{ij}$ is the fraction of samples in
$\mathcal C_{ij}$. Equation~(\ref{eq:pp_source_state_decomposition})
identifies how different source amplitude--angle populations contribute
to the measured third-order moment. Reconstructing the unconditioned
moment verifies the conditional accounting; it is not a separate test
of the PP relation.

Using the same midpoints and separation directions, we measure the
angle at smaller separations $r_<$. For every source-state cell, we
calculate
$\langle\theta_<-\theta_>\mid A_>,\theta_>\rangle$
and
$\langle(\theta_<-\theta_>)^2\mid A_>,\theta_>\rangle$.
Every ordered pair of the selected scales satisfying $r_<<r_>$ is
included, rather than selecting a fixed ratio between the two
separations.

The conditional PP moment and the angular-transition moments measure different quantities. The former determines how a specified source population contributes to
the third-order moment associated with transfer toward smaller scales.
The angular-transition moments determine how much the mutual angle of
that population changes. The PP relation provides no identity between
these quantities. Their joint evaluation identifies whether source populations contributing to the PP moment rotate toward smaller angle or undergo comparatively small angular changes.

The unconditioned third-order moment has the negative sign required by
Eq.~(\ref{eq:pp_perpendicular}) in $2654/2720$, or $97.6\%$, of the
separate separation--direction--octant estimates across the five stored
states and the two Els\"asser relations. Among the ten
stored-state--Els\"asser combinations, the coefficient of variation
with separation of the normalized quantity on the left-hand side of
Eq.~(\ref{eq:pp_scale_estimate}) ranges from $0.040$ to $0.251$.
The two perpendicular directions give the same sign in nearly all
cases, although their measured magnitudes are not equally well
converged in every stored state. The calculation reproduces the expected sign and approximate proportionality
to separation but does not provide a high-precision measurement of
$\varepsilon^\pm$.

The population-weighted conditional means reproduce the independently
measured PP moment with a largest relative difference of
$1.13\times10^{-15}$. The mean of the $+$ and $-$ conditional PP
moments for the largest-angle, highest-amplitude population is negative
at all nine source scales, in all five stored states, and for all three
state-space resolutions, giving $135/135$ such tests. For the same
five stored states, the highest-amplitude range has smaller
$M_\theta^{(2)}$ than the lowest-amplitude range in $3218/3240$, or
$99.3\%$, of the scale-pair--angle comparisons. Large-angle, high-amplitude populations contribute with the forward-transfer
sign while showing comparatively small angular changes.

\subsection{Physical-time angular persistence and advective-term attribution}
\label{subsec:temporal_nonlinear}

The time-resolved full-MHD data test the amplitude dependence at fixed
separation.
The local magnetic field, perpendicular plane, sampling direction, and
increment endpoints are reconstructed independently at each stored
time. We compare consecutive fields at the same midpoint and separation;
the construction is Eulerian and does not follow material structures.

Because the local perpendicular geometry changes between stored times,
we measure the relation between initial amplitude and subsequent angular
change after accounting for the initial angular state, the rotation of
the local sampling direction, and the change of the local perpendicular
plane. We quantify this amplitude dependence using the partial Spearman rank
correlations
\begin{align}\nonumber
&\rho_s
=
\rho_S\!\left[
\ln A_0,\,
|s_1-s_0|
\,\middle|\,
s_0,\alpha_r,\alpha_b
\right],\\\nonumber
&\,\rho_\theta
=
\rho_S\!\left[
\ln A_0,\,
|\theta_1-\theta_0|
\,\middle|\,
\theta_0,\alpha_r,\alpha_b
\right].
\end{align}
Here $\alpha_r$ is the change of the local perpendicular sampling
direction between the two times and $\alpha_b$ is the change of the
local mean-field direction, and hence of the local perpendicular plane.
The coefficients $\rho_s$ and $\rho_\theta$ control for the initial angular
state and changes of sampling geometry. The ordering also survives explicit
conditioning:
after dividing the source population into six initial-angle ranges,
the amplitude--angular-displacement correlation remains negative in
all 1536/1536 separation--time--angle tests for both $s$ and $\theta$;
details are given in Appendix~\ref{app:temporal_nonlinear}. Negative
values mean that larger-amplitude increment pairs undergo smaller
subsequent angular changes; see Table~\ref{tab:temporal_angular_persistence}.

\begin{table}
\caption{
Median interval-by-interval partial Spearman correlations
$\rho_s$ and $\rho_\theta$ between initial amplitude and subsequent
changes of $s=\sin\theta$ and of the folded angle $\theta$,
respectively, in the full-MHD simulation. The initial angular state and changes of the local sampling geometry
are controlled for. Both
correlations are negative in all sixty-four stored-time intervals at
every separation.
}
\label{tab:temporal_angular_persistence}
\centering
\begin{tabular*}{0.82\columnwidth}{@{\extracolsep{\fill}}c|cc@{}}
\hline
$r$ & $\rho_s$ & $\rho_\theta$ \\
\hline
$64$  & $-0.346$ & $-0.357$ \\
$80$  & $-0.340$ & $-0.352$ \\
$96$  & $-0.327$ & $-0.343$ \\
$112$ & $-0.318$ & $-0.334$ \\
\hline
\end{tabular*}
\end{table}

The same geometry applies in physical time: for a given vector change, a
larger increment turns through a smaller angle.

\begin{figure}
\centering
\includegraphics[width=0.9\textwidth]
{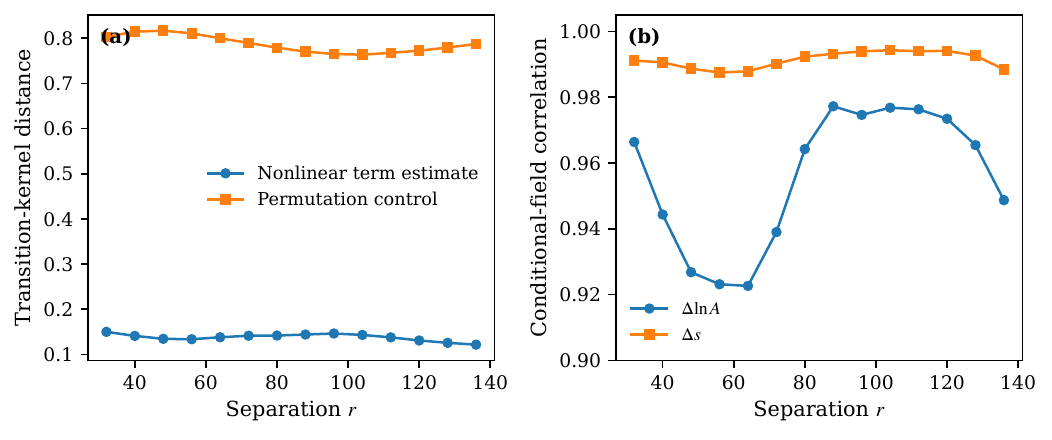}
\caption{
Multiscale finite-time attribution of the Els\"asser advective term in
the full-MHD simulation.
(a) Transition-kernel distance between the measured fixed-endpoint
evolution and the advective-term estimate, compared with the
permutation control.
(b) Conditional-field correlation between the measured and
advective-estimated changes of $\ln A$ and $s$. The calculation uses
the same finite-time construction at all fourteen separations.
}
\label{fig:multiscale_nonlinear_attribution}
\end{figure}
Both correlations are negative in every one of the sixty-four
time intervals at all four separations. Large-amplitude
Els\"asser-increment pairs undergo smaller subsequent changes
of their mutual angle at fixed separation, after the initial angular
state and motion of the local perpendicular geometry have been
accounted for. The result is obtained independently with $s$ and with
the folded angle itself. A spatially disjoint region gives the same
sign throughout the temporal record, and temporal transition
probabilities estimated in one region retain predictive information
when transferred to the second region and to later times excluded from
their estimation. Further details are given in
Appendix~\ref{app:temporal_nonlinear}.

To compare the measured change with the Els\"asser advective term, we
evaluate the finite-time change in a fixed source-time perpendicular
plane. For this calculation the
source-time endpoints and perpendicular plane are retained at the
destination time. The local Els\"asser advective terms measured from
the actual fields at the beginning and end of each stored-time interval
are combined to obtain the advective-term estimate. This is a
finite-time attribution along the measured MHD evolution rather than a
forward integration using that term alone. The calculation is then repeated over fourteen separations,
$r=32,40,\ldots,136$. Figure~\ref{fig:multiscale_nonlinear_attribution}
summarizes the scale dependence of the continuous attribution. The transition-kernel distance between the measured evolution and the
advective-term estimate remains between $0.122$ and $0.150$ over the
full separation range. The corresponding permutation control lies
between $0.764$ and $0.817$. The advective estimate is much closer to the measured source--destination
transition structure at every tested separation, with no systematic trend
across the range.

The conditional-field correlation for $\Delta\ln A$ ranges from
$0.923$ to $0.977$, while that for $\Delta s$ ranges from $0.988$ to
$0.994$. The finite-time advective update closely tracks the resolved
state dependence of both angular and amplitude changes across the tested
range. At fixed Eulerian points, however, the advective term contains
sweeping by the large-scale velocity and propagation along the large-scale
magnetic field, as well as scale-local distortion. These correlations do
not isolate the scale-local cascade interaction. A separate two-class
comparison uses large-amplitude small-angle and large-amplitude large-angle
populations. Once their source-amplitude distributions are
matched, the difference between their departure probabilities changes
sign through the temporal record rather than retaining a universal
ordering. The conditional-state comparison instead shows joint dependence on amplitude
and angle.

The destination-time advective term is evaluated from the actual evolved
field, and the difference between the measured and estimated changes also
contains contributions from pressure, dissipation, forcing, finite-time
quadrature, spatial differentiation, and interpolation. The update retains a
strong dependence on the initial amplitude--angle state but is not an
autonomous evolution model.

\section{Discussion and conclusions}
\label{sec:conclusions}

At fixed initial angle, large-amplitude Els\"asser-increment pairs undergo
smaller angular changes across scale in both full MHD and RMHD, including
states that begin at large angles. A fixed vector change turns a larger
increment through a smaller angle, so the ordering contains a geometric
contribution. Signed drift and displacement magnitude remain distinct: a mean
shift toward smaller angle can coexist with amplitude-dependent angular
persistence.

The ordering occurs in all 132 JHTDB starting-scale--angle combinations, in
658/660 and 659/660 adjacent-amplitude comparisons, and in the RMHD direction,
octant, and second-state tests. In physical time, the amplitude--displacement
correlations are negative in all 1536 conditional tests after controlling for
initial angle and changes of local geometry.

The one-scale moments $I_2$ and $I_3$ increase toward smaller separation, and
stronger amplitude weighting gives smaller $\langle s_r\rangle_{A^p}$ with
stronger scale dependence. In the 50 scale transitions with a decreasing
conventional weighted statistic, neither angular change at fixed amplitude
weights nor amplitude reweighting at fixed angle produces the decrease; it
requires their joint change. The Politano--Pouquet analysis gives a separate
result: large-angle, high-amplitude source populations contribute with the
forward-transfer sign while showing comparatively small angular displacement.
The PP relation itself does not determine mutual-angle evolution because its
longitudinal increment is defined relative to the separation direction.

At fixed separation, the same amplitude ordering appears in physical time,
with the same geometric caveat. A finite-time update based on the Els\"asser
advective term tracks the conditional amplitude--angle dependence. At fixed
Eulerian points, this term includes large-scale sweeping and propagation along
the large-scale magnetic field as well as scale-local distortion, so the
correlations do not isolate scale-local cascade transfer. Pressure,
dissipation, forcing, and the remaining MHD terms also contribute.

The Chapman--Kolmogorov discrepancy is smaller than the variation between
spatial half-volumes but larger than sampling noise alone; the reduced state
$(\ln A,s)$ is approximately, not exactly, Markov at the tested resolution.
Kernels estimated in one subvolume retain predictive information in a disjoint
subvolume and, in the temporal test, at held-out later intervals. These tests
use one DNS realization and do not establish nonlinear locality or
universality. The scale sequences are matched Eulerian measurements rather
than trajectories of coherent structures, and both simulations cover finite
scale ranges.

The conventional amplitude-weighted alignment diagnostic is one projection of
the joint state distribution and transition law. Its scale dependence does not
by itself specify the signed angular drift of the increment population.

\section*{Data availability}

The full-MHD data used in this work are publicly available through the
Johns Hopkins Turbulence Database (JHTDB)
\citep{JHTB1,JHTB2,Eyinketal2013}. The analysis uses the \texttt{mhd1024} data set,
with the spatial regions, stored times, separations, and sampling
procedures specified in the appendices.

The reduced-MHD data are from the high-resolution strong-guide-field
simulations described by \citet{Beresnyak2014,Beresnyak2015} and are
publicly available in the ``Shared MHD datacubes'' collection hosted on
Globus (collection ID
\texttt{cd71b5d6-8850-426a-b057-0b269b10fb82}). The present analysis uses the ordinary-viscosity \texttt{b1024n} run.
The angular-persistence calculation uses \texttt{R108000} as its
primary stored state and \texttt{R162000} for its original
reproducibility test. The higher-order one-scale statistics and the Politano--Pouquet
calculation use the five stored states \texttt{R018000},
\texttt{R072000}, \texttt{R108000}, \texttt{R126000}, and
\texttt{R162000}. Further details of these calculations are given in
Appendix~\ref{app:beresnyak}.

The analysis and figure-generation scripts used for the results reported
in this paper are available at
\url{https://zenodo.org/records/22218391}.

\appendix

\section{Stochastic evolution in scale}
\label{app:scale_stochastic}

Section~\ref{sec:stochastic_formalism} uses a finite-step stochastic description of the reduced state $(a,s)$. This appendix gives the corresponding two-scale joint probability, conditional kernels, multiscale composition, continuous path representation, general Markov generators and Kramers--Moyal expansions, and the projection onto the conventional amplitude-weighted alignment measure. Throughout this appendix,
$\boldsymbol{\xi}=(a,s)$ with $a=\ln A\in\mathbb{R}$ and
$0\leq s\leq1$, and
$d\boldsymbol{\xi}=da\,ds$. The logarithmic scale coordinate
$\tau=\ln(r_0/r)$ increases toward smaller physical separation.


For two ordered scales $\tau_2>\tau_1$, the complete two-scale
description of the reduced amplitude--angle state is the joint
probability density $
p(\boldsymbol{\xi}_2,\tau_2;
  \boldsymbol{\xi}_1,\tau_1)$. The one-scale probability densities are obtained as its two
marginals,
\begin{align}
p(\boldsymbol{\xi}_1,\tau_1)
&=
\int
p(\boldsymbol{\xi}_2,\tau_2;
  \boldsymbol{\xi}_1,\tau_1)
\,d\boldsymbol{\xi}_2 ,
\nonumber\\
p(\boldsymbol{\xi}_2,\tau_2)
&=
\int
p(\boldsymbol{\xi}_2,\tau_2;
  \boldsymbol{\xi}_1,\tau_1)
\,d\boldsymbol{\xi}_1 .
\label{eq:app_joint_marginals}
\end{align}
Conditioning on the larger-scale state defines the transition kernel
toward smaller separation,
\begin{align}
K_{2|1}
&\equiv
K(\boldsymbol{\xi}_2,\tau_2
  \mid\boldsymbol{\xi}_1,\tau_1)
\nonumber\\
&=
\frac{
p(\boldsymbol{\xi}_2,\tau_2;
  \boldsymbol{\xi}_1,\tau_1)
}{
p(\boldsymbol{\xi}_1,\tau_1)
}.
\label{eq:app_forward_kernel}
\end{align}
For every allowed source state this kernel is normalized,
\begin{align}
\int
K(\boldsymbol{\xi}_2,\tau_2
  \mid\boldsymbol{\xi}_1,\tau_1)
\,d\boldsymbol{\xi}_2
&=1.
\label{eq:app_kernel_normalization}
\end{align}
The same joint probability can instead be conditioned on the
smaller-scale state,
\begin{align}
K_{1|2}
&\equiv
K(\boldsymbol{\xi}_1,\tau_1
  \mid\boldsymbol{\xi}_2,\tau_2)
\nonumber\\
&=
\frac{
p(\boldsymbol{\xi}_2,\tau_2;
  \boldsymbol{\xi}_1,\tau_1)
}{
p(\boldsymbol{\xi}_2,\tau_2)
}.
\label{eq:app_reverse_kernel}
\end{align}
The two conditional descriptions are related by
\begin{align}
K_{1|2}
&=
K_{2|1}
\frac{
p(\boldsymbol{\xi}_1,\tau_1)
}{
p(\boldsymbol{\xi}_2,\tau_2)
}.
\label{eq:app_kernel_bayes}
\end{align}
Thus the reverse conditional probability is not obtained, in general,
by simply exchanging the arguments of the forward kernel. The two
kernels answer different conditional questions but reconstruct the same
joint probability.

The terminology ``forward'' and ``backward'' requires some care.
Equation~(\ref{eq:app_kernel_bayes}) concerns conditioning in opposite
directions through scale. This is distinct from the forward and
backward Kolmogorov equations. For a continuous Markov process with
generator ${\cal L}_\tau$, the same forward transition probability
satisfies schematically
\begin{align}
\partial_{\tau_2}K_{2|1}
&=
{\cal L}_{\tau_2}^{\dagger}K_{2|1},
\nonumber\\
-\partial_{\tau_1}K_{2|1}
&=
{\cal L}_{\tau_1}K_{2|1}.
\label{eq:app_forward_backward_kolmogorov}
\end{align}
Here ``forward'' or ``backward'' refers to whether the evolution
operator acts on the destination or source variables, not to replacing
$K_{2|1}$ by the Bayes-reversed kernel $K_{1|2}$.

Integrating the joint probability over the source state gives the
propagation law used in the main text,
\begin{align}
p(\boldsymbol{\xi}_2,\tau_2)
&=
\int
K(\boldsymbol{\xi}_2,\tau_2
  \mid\boldsymbol{\xi}_1,\tau_1)
p(\boldsymbol{\xi}_1,\tau_1)
\,d\boldsymbol{\xi}_1 .
\label{eq:app_probability_propagation}
\end{align}
No Markov assumption is required for
Eqs.~(\ref{eq:app_forward_kernel})--(\ref{eq:app_probability_propagation});
they are exact relations for any pair of scales.

\subsection{Composition through many scales}
\label{app:multiscale_composition}

The Markov assumption enters only when transitions through several
scales are constructed from neighbouring-scale transitions. For
$\tau_1<\tau_2<\tau_3$, it requires
\begin{align}
p(\boldsymbol{\xi}_3
  \mid\boldsymbol{\xi}_2,\boldsymbol{\xi}_1)
&=
p(\boldsymbol{\xi}_3
  \mid\boldsymbol{\xi}_2),
\label{eq:app_markov_property}
\end{align}
which gives the Chapman--Kolmogorov relation
\begin{align}
&K(\boldsymbol{\xi}_3,\tau_3
   \mid\boldsymbol{\xi}_1,\tau_1)
\nonumber\\
&\quad =
\int
K(\boldsymbol{\xi}_3,\tau_3
  \mid\boldsymbol{\xi}_2,\tau_2)
K(\boldsymbol{\xi}_2,\tau_2
  \mid\boldsymbol{\xi}_1,\tau_1)
\,d\boldsymbol{\xi}_2 .
\label{eq:app_chapman_kolmogorov}
\end{align}
The numerical test of this relation is described separately in
Appendix~\ref{app:sampling_ck}; here we use it only as the condition
that permits successive scale transitions to be composed.

For a sequence
$\tau_0<\tau_1<\cdots<\tau_N$, the joint probability of a complete
scale history factorizes as
\begin{align}
&p(\boldsymbol{\xi}_0,\ldots,\boldsymbol{\xi}_N)
\nonumber\\
&\quad =
p(\boldsymbol{\xi}_0,\tau_0)
\prod_{j=0}^{N-1}
K(\boldsymbol{\xi}_{j+1},\tau_{j+1}
  \mid\boldsymbol{\xi}_j,\tau_j).
\label{eq:app_discrete_path_probability}
\end{align}
The endpoint transition probability follows by integrating over all
intermediate states,
\begin{align}
&K(\boldsymbol{\xi}_N,\tau_N
   \mid\boldsymbol{\xi}_0,\tau_0)
\nonumber\\
&=
\int
\left[
\prod_{j=1}^{N-1}d\boldsymbol{\xi}_j
\right]
\prod_{j=0}^{N-1}
K(\boldsymbol{\xi}_{j+1},\tau_{j+1}
  \mid\boldsymbol{\xi}_j,\tau_j).
\label{eq:app_path_composition}
\end{align}
Equation~(\ref{eq:app_path_composition}) is the finite-step path
representation used in Section~\ref{sec:stochastic_formalism}. Each
possible sequence of intermediate amplitude--angle states contributes
with the product of its neighbouring-scale transition probabilities.

The transition process need not be homogeneous in scale. The kernel
can depend explicitly on $\tau$, or equivalently on the physical
separation $r=r_0e^{-\tau}$. Propagation over a finite interval
therefore requires an ordered product of the appropriate kernels and
cannot in general be replaced by a power of one scale-independent
transition operator.

The logarithmic variable is useful because physical scale ratios then
become additive intervals,
\begin{align}
\tau_2-\tau_1
&=
\ln\left(\frac{r_1}{r_2}\right).
\label{eq:app_log_scale_interval}
\end{align}
A small scale-space step $\Delta\tau\rightarrow0$ corresponds to a small fractional change $|\Delta r|/r\rightarrow0$; the physical separation need not approach the dissipation range. The physical values of $r$ determine whether a scale history lies in the inertial or dissipation range.

\subsection{Continuous scale generator and functional representation}
\label{app:continuous_scale_generator}

The finite-step formulation does not require an infinitesimal model.
Nevertheless, if a controlled continuous limit exists, the
neighbouring-scale kernels can be represented by a scale-evolution
generator.

Let
$\Delta\boldsymbol{\xi}
=\boldsymbol{\xi}'-\boldsymbol{\xi}$
be the change of state over a short interval $\Delta\tau$. We first
describe the complete short-step distribution through its conditional
characteristic function,
\begin{align}
\Phi(\boldsymbol{\pi};
     \boldsymbol{\xi},\tau,\Delta\tau)
&=
\int
e^{i\boldsymbol{\pi}\cdot\Delta\boldsymbol{\xi}}
\nonumber\\
&\quad\times
K(\boldsymbol{\xi}+\Delta\boldsymbol{\xi},
  \tau+\Delta\tau
  \mid\boldsymbol{\xi},\tau)
\,d\Delta\boldsymbol{\xi}.
\label{eq:app_characteristic_function}
\end{align}
The auxiliary variable $\boldsymbol{\pi}$ is conjugate to the two
components $(a,s)$ of the reduced state; it is unrelated to an MHD
wavevector.

If a continuous generator exists, its local characteristic symbol is
defined by
\begin{align}
\Psi(\boldsymbol{\xi},\boldsymbol{\pi},\tau)
&=
\lim_{\Delta\tau\rightarrow0}
\frac{
\ln\Phi(
\boldsymbol{\pi};
\boldsymbol{\xi},\tau,\Delta\tau)
}{
\Delta\tau
}.
\label{eq:app_generator_symbol}
\end{align}
For a sufficiently short interval,
\begin{align}
\Phi
&=
\exp\left[
\Delta\tau\,
\Psi(
\boldsymbol{\xi},\boldsymbol{\pi},\tau)
+o(\Delta\tau)
\right].
\label{eq:app_characteristic_short_step}
\end{align}
Fourier inversion then gives the short-step kernel in the form
\begin{align}
&K(\boldsymbol{\xi}_{j+1},\tau_{j+1}
   \mid\boldsymbol{\xi}_j,\tau_j)
\nonumber\\
&\simeq
\int
\frac{d\boldsymbol{\pi}_j}{(2\pi)^2}
\exp\bigg[
-i\boldsymbol{\pi}_j\cdot
(\boldsymbol{\xi}_{j+1}-\boldsymbol{\xi}_j)
\nonumber\\
&\hspace{29mm}
+\Delta\tau\,
\Psi(
\boldsymbol{\xi}_j,
\boldsymbol{\pi}_j,\tau_j)
\bigg].
\label{eq:app_short_step_fourier}
\end{align}
Substituting this expression into
Eq.~(\ref{eq:app_path_composition}) and taking the continuum limit
gives the formal functional representation
\begin{align}
&K(\boldsymbol{\xi}_f,\tau_f
   \mid\boldsymbol{\xi}_i,\tau_i)
\nonumber\\
&=
\int
{\cal D}\boldsymbol{\xi}\,
{\cal D}\boldsymbol{\pi}\,
\exp\left\{
\int_{\tau_i}^{\tau_f}
d\tau\,
\left[
-i\boldsymbol{\pi}\cdot
\dot{\boldsymbol{\xi}}
+
\Psi(
\boldsymbol{\xi},
\boldsymbol{\pi},\tau)
\right]
\right\}.
\label{eq:app_functional_integral}
\end{align}
Equation~(\ref{eq:app_functional_integral}) is the continuous
counterpart of summing over all intermediate amplitude--angle states.
The expression is formal: its detailed normalization and the precise
form of the functional measure depend on the chosen discretization of
the short-step process. The finite-step transition law
Eq.~(\ref{eq:app_path_composition}) is the underlying quantity and does
not depend on this representation.

The generator
$\Psi$ has not been assumed to describe ordinary diffusion. It can
contain higher-order or nonlocal changes of the amplitude--angle state.
A Fokker--Planck or Langevin model corresponds only to a restricted
class of generators.

\subsection{Kramers--Moyal expansion and the diffusion limit}
\label{app:kramers_moyal}

When the short-step generator admits an expansion in conditional
moments, define the Kramers--Moyal coefficients by
\begin{align}
D^{(n)}_{i_1\cdots i_n}
(\boldsymbol{\xi},\tau)
&=
\frac{1}{n!}
\lim_{\Delta\tau\rightarrow0}
\frac{1}{\Delta\tau}
\nonumber\\
&\quad\times
\left\langle
\Delta\xi_{i_1}\cdots
\Delta\xi_{i_n}
\mid
\boldsymbol{\xi},\tau
\right\rangle .
\label{eq:app_km_coefficients}
\end{align}
The one-scale probability density then obeys
\begin{align}
\partial_\tau p
&=
\sum_{n=1}^{\infty}
(-1)^n
\partial_{i_1}\cdots\partial_{i_n}
\left[
D^{(n)}_{i_1\cdots i_n}p
\right],
\label{eq:app_km_hierarchy}
\end{align}
where repeated state indices are summed over the two coordinates
$(a,s)$. The same coefficients appear in the expansion of the
generator symbol,
\begin{align}
\Psi(
\boldsymbol{\xi},\boldsymbol{\pi},\tau)
&=
\sum_{n=1}^{\infty}
D^{(n)}_{i_1\cdots i_n}\times
(i\pi_{i_1})\cdots(i\pi_{i_n}).
\label{eq:app_symbol_km}
\end{align}
If only the first-order term is present, the probability distribution
is transported deterministically through state space, giving a
Liouville-type evolution. If the hierarchy terminates after second
order, the result is a Fokker--Planck equation and can, under the usual
conditions, be represented by a Langevin process.

A finite truncation at an order higher than two is not generally consistent with a positive Markov probability density. Pawula's theorem requires the Kramers--Moyal hierarchy either to terminate after the second term or to continue to all orders. A genuine nonzero fourth-order coefficient therefore rules out a consistent fourth-order truncation. The theorem does not establish that every higher-order coefficient is nonzero; it rules out retaining only a finite number of terms beyond second order.

The Kramers--Moyal hierarchy is itself not the most general possible
Markov generator. Processes containing nonlocal changes in state space
can lead naturally to integro-differential generators for which a
derivative expansion is not useful. The finite-step transition-kernel
formulation is therefore more general than either a Fokker--Planck
equation or a truncated Kramers--Moyal model.

\subsection{Observable averages and the weighted alignment measure}
\label{app:observable_projection}

The joint probability also gives the scale evolution of any observable
$F(\boldsymbol{\xi})$. At the smaller scale,
\begin{align}
\langle F\rangle_2
&=
\iint
F(\boldsymbol{\xi}_2)
p(\boldsymbol{\xi}_2,\tau_2;
  \boldsymbol{\xi}_1,\tau_1)
\,d\boldsymbol{\xi}_1
\,d\boldsymbol{\xi}_2 .
\end{align}
Writing the corresponding expression at the larger scale and
subtracting gives
\begin{align}
\langle F\rangle_2-\langle F\rangle_1
&=
\iint
\bigl[
F(\boldsymbol{\xi}_2)
-
F(\boldsymbol{\xi}_1)
\bigr]
\nonumber\\
&\quad\times
p(\boldsymbol{\xi}_2,\tau_2;
  \boldsymbol{\xi}_1,\tau_1)
\,d\boldsymbol{\xi}_1
\,d\boldsymbol{\xi}_2 .
\label{eq:app_observable_change}
\end{align}
Using the forward conditional kernel, this may also be written as the
larger-scale population average of the conditional changes,
\begin{align}
\langle F\rangle_2-\langle F\rangle_1
&=
\int
p(\boldsymbol{\xi}_1,\tau_1)
\nonumber\\
&\quad\times
\left\langle
F_2-F_1
\mid
\boldsymbol{\xi}_1
\right\rangle
d\boldsymbol{\xi}_1 .
\label{eq:app_observable_conditional}
\end{align}
Equations~(\ref{eq:app_observable_change}) and
(\ref{eq:app_observable_conditional}) make explicit that the scale
dependence of a global statistic is obtained by averaging the
conditional changes of the different amplitude--angle states.

The conventional sine-based alignment measure is one particular
projection of the one-scale distribution. Since $A=e^a$,
\begin{align}
\langle s\rangle_A(\tau)
&=
\frac{
\displaystyle
\int e^a s\,
p(a,s;\tau)\,da\,ds
}{
\displaystyle
\int e^a
p(a,s;\tau)\,da\,ds
}.
\label{eq:app_weighted_projection}
\end{align}
After propagation from $\tau_1$ to $\tau_2$, use of
Eq.~(\ref{eq:app_probability_propagation}) gives
\begin{align}
\langle s\rangle_A(\tau_2)
&=
\frac{
\displaystyle
\iint
e^{a_2}s_2
K_{2|1}
p(\boldsymbol{\xi}_1,\tau_1)
\,d\boldsymbol{\xi}_1d\boldsymbol{\xi}_2
}{
\displaystyle
\iint
e^{a_2}
K_{2|1}
p(\boldsymbol{\xi}_1,\tau_1)
\,d\boldsymbol{\xi}_1d\boldsymbol{\xi}_2
},
\label{eq:app_weighted_kernel_projection}
\end{align}
where
$K_{2|1}=
K(\boldsymbol{\xi}_2,\tau_2
\mid\boldsymbol{\xi}_1,\tau_1)$.
Thus the conventional weighted alignment measure depends on the full
joint transition of amplitude and angle.

A more direct physical decomposition follows by writing
$A_2=A_1+\Delta A$ and $s_2=s_1+\Delta s$. Starting from
$\langle s\rangle_{A,2}
=\langle A_2s_2\rangle/\langle A_2\rangle$ and subtracting
$\langle s\rangle_{A,1}$ gives exactly
\begin{align}
&\langle s\rangle_{A,2}
-
\langle s\rangle_{A,1}
\nonumber\\
&=
\frac{1}{\langle A_2\rangle}
\bigg[
\left\langle A_1\Delta s\right\rangle
\nonumber\\
&\qquad+
\left\langle
\bigl(
s_1-\langle s\rangle_{A,1}
\bigr)
\Delta A
\right\rangle
+
\left\langle
\Delta A\,\Delta s
\right\rangle
\bigg].
\label{eq:app_weighted_finite_step_decomposition}
\end{align}
The first term measures angular change weighted by the amplitude
already present at the larger scale. The second measures changes of
amplitude that depend on the initial angular state. The third measures
the part of the evolution in which amplitude and angle change together
during the same scale interval.

Equation~(\ref{eq:app_weighted_finite_step_decomposition}) is the
finite-step form used in the main text. It shows explicitly why the
scale dependence of the amplitude-weighted alignment measure cannot in
general be identified with angular rotation alone. The weighted
diagnostic is determined by the complete joint redistribution of
amplitude and angle.

\section{Full-MHD scale-space sampling and numerical consistency tests}
\label{app:sampling_ck}

This appendix gives the numerical details of the matched scale-space
measurements, the one-scale intermittency statistics, the finite-step
decomposition of the weighted statistic, the Chapman--Kolmogorov test,
and the out-of-sample spatial transfer used for the full-MHD results
in Section~\ref{subsec:scale_tests}.

\subsection{Matched scale-space measurements}
\label{app:matched_scale_measurements}

The scale-space calculation uses nine stored fields with indices
$1$, $129$, $257$, $385$, $513$, $641$, $769$, $897$, and $1025$.
The primary sample, denoted below as Cube A, is a fixed $448^3$
subvolume with one-based grid coordinates
$x,y,z=289,\ldots,736$. The same $4000$ spatial midpoints are used in
each stored field. At every midpoint and separation, eight unoriented
sampling axes are constructed in the plane perpendicular to the local
mean magnetic field, giving $32000$ midpoint--direction measurements
per stored field and separation. The stored fields provide repeated
samples of the same spatial volume rather than independent spatial
realizations.

The local mean magnetic field is recalculated at each separation using
a three-dimensional Gaussian filter with standard deviation $r/2$ in
full-grid units. For this operation the magnetic field is first block
averaged by a factor of four, so the Gaussian standard deviation on the
reduced grid is $r/8$; the kernel is truncated at three standard
deviations.

A fixed reference vector $\mathbf e_{\rm ref}$ is assigned to each
midpoint and used to label the perpendicular sampling axes consistently
as the local magnetic geometry changes with separation. With
$\widehat{\mathbf b}_r$ denoting the local mean-field direction, the
basis, sampling axes, and centered endpoints are
\begin{align}
\mathbf e_1(r)
&=
\frac{
\mathbf e_{\rm ref}
-
(\mathbf e_{\rm ref}\cdot\widehat{\mathbf b}_r)
\widehat{\mathbf b}_r
}{
\left|
\mathbf e_{\rm ref}
-
(\mathbf e_{\rm ref}\cdot\widehat{\mathbf b}_r)
\widehat{\mathbf b}_r
\right|
},
\qquad
\mathbf e_2(r)
=
\widehat{\mathbf b}_r\times\mathbf e_1(r),
\nonumber\\
\widehat{\mathbf r}_j(r)
&=
\cos\!\left(\frac{\pi j}{8}\right)\mathbf e_1(r)
+
\sin\!\left(\frac{\pi j}{8}\right)\mathbf e_2(r),
\,
j=0,\ldots,7,
\nonumber\\
\mathbf x_\pm(r,j)
&=
\mathbf x
\pm
\frac{r}{2}\widehat{\mathbf r}_j(r).
\label{eq:appB_sampling_geometry}
\end{align}
The magnitude of the projected reference vector remains larger than
$0.2000$ for every retained measurement, so this construction remains
well conditioned. Velocity and magnetic fields at the generally
off-grid endpoints are evaluated by trilinear interpolation. The
centered Els\"asser increments are then projected into the local
perpendicular plane as in the main text.

The calculation uses $25$ logarithmically spaced separations from
$r=96$ to $r=32$, with elementary spacing
$\delta\tau=\ln(3)/24=0.045775512$. A scale sequence keeps the same
stored field, midpoint, and direction label while the local mean
magnetic field, perpendicular plane, physical sampling direction, and
increment endpoints are reconstructed at every separation. Thus the
scale sequence consists of matched equal-time spatial measurements and
does not represent the trajectory of an identifiable turbulent
structure.

For the angular-persistence calculation, the measurements at each
starting separation are first divided into six equal-population
ranges in $s$.  Within each initial-angle range, the source
amplitudes $a=\ln A$ are then divided into six equal-population
ranges.  This gives six amplitude populations at each of six
initial-angle ranges.  The scale pair is separated by three entries of
the logarithmic grid, leaving 22 possible starting separations.  The
smallest- and largest-amplitude ranges at each fixed initial-angle
range give the small- and large-amplitude curves in Fig.~\ref{fig:scale_space_summary}.

For each of the resulting $22\times6=132$
starting-scale--initial-angle combinations, the dependence of
$M_s^{(2)}$ and $M_\theta^{(2)}$ on the six amplitude ranges was
tested in three complementary ways.  First, the Spearman rank correlation between the median source
amplitudes of the six amplitude ranges and the angular displacement is
negative in all 132 cases for both angular measures.
Second, the largest-amplitude range has smaller displacement than
the smallest-amplitude range in all 132/132 cases for both
$M_s^{(2)}$ and $M_\theta^{(2)}$.  Third, considering every pair of
successive amplitude ranges separately gives 660 adjacent-amplitude
comparisons: the displacement decreases in 658/660 comparisons for
$M_s^{(2)}$ and 659/660 for $M_\theta^{(2)}$.  The median
largest-to-smallest displacement ratios are 0.494 and 0.497,
respectively.  The reported persistence is not inferred from
the two extreme amplitude populations alone and is nearly monotonic
through the complete six-range amplitude ordering. The representative transition
$r_>=63.5844\rightarrow r_<=55.4256$ is one member of this sequence
and is not selected by its measured persistence.

\subsection{One-scale intermittency statistics}
\label{app:full_mhd_higher_moments}

The full-MHD one-scale statistics shown in
Fig.~\ref{fig:higher_moment_alignment} are evaluated separately from
the matched scale-transition calculation above. We use fifteen
$320^3$ subvolumes and the nine separations
\[
r=32,\ 40,\ 48,\ 64,\ 80,\ 96,\ 128,\ 160,\ 192.
\]
For each subvolume, $120000$ spatial midpoints are sampled with eight
local-perpendicular direction labels at $45^\circ$ intervals. Centered
increments are unchanged under direction reversal, so these labels represent
four independent axes counted twice: $960000$ evaluated direction labels, or
$480000$ independent midpoint--axis samples, per subvolume and separation. The local mean magnetic field is constructed
with the same Gaussian width and block-averaging procedure described
above, and the Els\"asser increments are projected into the resulting
local perpendicular plane before $A_r$ and $s_r$ are evaluated.

For each subvolume and separation we calculate $I_2(r)$, $I_3(r)$,
and $\langle s_r\rangle_{A^p}$ for $p=0,1,2,3$. The values shown in
Fig.~\ref{fig:higher_moment_alignment} are the means across the
fifteen subvolumes, and the error bars are the corresponding standard
errors. The individual spatial measurements within a subvolume are
therefore not treated as independent ensemble realizations. The strict
ordering
\[
\langle s_r\rangle_{A^0}>
\langle s_r\rangle_A>
\langle s_r\rangle_{A^2}>
\langle s_r\rangle_{A^3}
\]
holds in all $135$ subvolume--separation combinations.

\subsection{Finite-step decomposition of the weighted statistic}

We also evaluate the three terms in Eq.~(\ref{eq:weighted_finite_step_decomposition}) directly from
the same matched projected-perpendicular scale sequences.
The calculation is performed independently in the primary
448$^3$ sample and in the spatially disjoint 448$^3$ sample
defined below.  For each scale transition we use the matched
values $(A_>,s_>)$ and $(A_<,s_<)$ of every retained
measurement and calculate
\[
C_{\rm ang}
=
\frac{\langle A_>\Delta s\rangle}{\langle A_<\rangle},
\qquad
C_{\rm amp}
=
\frac{
\langle
(s_>-\langle s\rangle_{A,>})\Delta A
\rangle
}{
\langle A_<\rangle
},
\]
and
\[
C_{\rm joint}
=
\frac{\langle\Delta A\,\Delta s\rangle}
{\langle A_<\rangle}.
\]
Their sum reproduces
$\langle s\rangle_{A,<}-\langle s\rangle_{A,>}$
to numerical precision.

For the three-grid-step transitions used in the
angular-persistence analysis, the weighted statistic decreases
in 19 of 22 transitions in the primary full-MHD sample and in
all 22 transitions in the spatially disjoint sample.  In all
41 of these decreasing transitions, $C_{\rm joint}<0$ and
$C_{\rm joint}$ is the largest contribution in absolute
magnitude.  Neither $C_{\rm ang}$ nor $C_{\rm amp}$ is
negative in any of the 41 transitions.  The median fractions
of
$|C_{\rm ang}|+|C_{\rm amp}|+|C_{\rm joint}|$
are approximately $0.22$, $0.23$, and $0.56$ in the primary
sample and $0.20$, $0.23$, and $0.57$ in the spatially
disjoint sample.

To test the spatial stability of this decomposition, every
stored field is also divided into eight spatial subvolumes.
For the direct transition from the maximum of the pooled
weighted statistic to the smallest sampled separation, the
joint term is the largest contribution in 67/72 spatial
blocks in the primary sample and 66/72 in the spatially
disjoint sample.  Its interquartile ranges are
$[-0.158,-0.101]$ and $[-0.141,-0.089]$, respectively,
whereas the corresponding interquartile ranges of the pure
amplitude-reweighting term are entirely positive,
$[0.027,0.057]$ and $[0.023,0.042]$.
These blockwise checks support the conclusion that the decrease requires
the joint amplitude--angle change and is not produced by pooling a small
number of exceptional regions.

\subsection{Finite-step consistency and spatial transfer}
\label{app:finite_step_transfer}

For the Chapman--Kolmogorov test, the joint $(a,s)$ state space is
partitioned independently at each separation into equal-population
ranges. The calculation is repeated with five, six, and seven ranges
in each variable, corresponding to $25$, $36$, and $49$ joint states.
Once determined, the boundaries at a given separation are held fixed
for every transition entering or leaving that separation.

For two scale indices $i<j$, let
$N_{\alpha\leftarrow\beta}^{\,i\rightarrow j}$ denote the number of
matched measurements in source state $\beta$ at $i$ and destination
state $\alpha$ at $j$. The empirical transition matrix is
\begin{align}
T_{\alpha\beta}^{\,i\rightarrow j}
&=
\frac{
N_{\alpha\leftarrow\beta}^{\,i\rightarrow j}
}{
\sum_\gamma
N_{\gamma\leftarrow\beta}^{\,i\rightarrow j}
},
\qquad
\sum_\alpha
T_{\alpha\beta}^{\,i\rightarrow j}
=
1 .
\label{eq:appB_transition_matrix}
\end{align}
For three equally spaced indices $i$, $i+g$, and $i+2g$, the transition
measured directly between the first and third scales is compared with
the transition composed through the intermediate scale. We quantify
their probability-distribution distance with the source-weighted total
variation
\begin{align}
D_{\rm direct\mbox{-}composed}(i,g)
&=
\frac{1}{2}
\sum_\beta p_\beta^{\,i}
\sum_\alpha
\Bigg|
T_{\alpha\beta}^{\,i\rightarrow i+2g}
\nonumber\\
&\qquad
-
\sum_\gamma
T_{\alpha\gamma}^{\,i+g\rightarrow i+2g}
T_{\gamma\beta}^{\,i\rightarrow i+g}
\Bigg|.
\label{eq:appB_direct_composed}
\end{align}
where $p_\beta^{\,i}$ is the empirical source-state probability. Here
$\Delta\tau=g\delta\tau$ is the interval between successive members of
the three-scale test; the direct first-to-third transition spans
$2\Delta\tau$. We evaluate $g=1,\ldots,8$.

The finite spatial precision of this comparison is measured directly
from the data. Each stored field is divided into eight spatial octants.
For each of $500$ random balanced divisions, four octants are assigned
to one spatial half-sample and the remaining four to the other.
Direct and composed transition matrices are then estimated separately
from the two halves. Using the same source weighting as in
Eq.~(\ref{eq:appB_direct_composed}), the spatial sampling variation is
\begin{align}
D_{\rm sampling}
&=
\frac{1}{2}
\left[
D_{\rm TV}
\left(
T_{\rm dir}^{(1)},T_{\rm dir}^{(2)}
\right)
+
D_{\rm TV}
\left(
T_{\rm comp}^{(1)},T_{\rm comp}^{(2)}
\right)
\right],
\label{eq:appB_sampling_variation}
\end{align}
where the superscripts denote the two disjoint half-samples and $D_{\rm TV}$ is the same source-weighted total-variation
distance between transition probabilities used in
Eq.~(\ref{eq:appB_direct_composed}). The
reported values are averaged over the random divisions and admissible
starting separations, with all nine stored fields contributing to the
comparison.

For all three state-space resolutions, the
starting-separation-averaged direct--composed difference is below the
half-volume variation at the finest tested interval and remains so over
the tested interval range. The measured ratios at the finest interval
are $0.91$, $0.81$, and $0.72$ for the $25$-, $36$-, and $49$-state
representations. The corresponding null expectation for the full-sample
discrepancy is about $0.45$, so the half-volume comparison is more lenient
than a sampling-noise test. The starting-separation audit in
Fig.~\ref{fig:finite_step_ck}(b) shows five marginal exceptions among the
$23$ starting separations for the $5\times5$ representation, with a
largest ratio of $1.074$. The departure from Chapman--Kolmogorov
composition is smaller than the variation between spatial half-volumes
but larger than sampling noise alone would give; the reduced state is
approximately, not exactly, Markov at this resolution.

We also test whether the measured scale-dependent transition law
retains predictive information outside the spatial region from which it
is estimated. A second $448^3$ subvolume, Cube B, is obtained by a
periodic half-box translation of Cube A. In each coordinate its wrapped
one-based range is
$801,\ldots,1024,1,\ldots,224$, so the two cubes contain no common
grid points. Cube B is sampled at the same nine stored fields, with the
same $25$ separations and the same numbers of midpoint and
local-perpendicular direction labels.

For this test, the six equal-population boundaries in $a$ and the six
in $s$ are determined from Cube A at each separation and then applied
without refitting to Cube B. Let $T_j^A$ denote the neighboring-scale
transition matrix measured in Cube A. Because the scale evolution is
inhomogeneous, predictions over several scale steps use the ordered
product
$T^A_{j,h}=T^A_{j+h-1}\cdots T^A_{j+1}T^A_j$, rather than a power of
one scale-independent matrix.

The conditional predictive content of this transferred kernel is
measured relative to the unconditional Cube-A destination
distribution. For a horizon of $h$ scale steps, the mean information
gain for the measured Cube-B transitions is
\begin{align}
G_h
&=
\frac{1}{N}
\sum_{\ell=1}^{N}
\log_2
\frac{
T^A_{j,h}
\left(
\alpha_\ell\mid\beta_\ell
\right)
}{
p^A_{j+h}(\alpha_\ell)
},
\label{eq:appB_transfer_gain}
\end{align}
where $\beta_\ell$ and $\alpha_\ell$ are the measured Cube-B source
and destination states. A positive value means that retaining the
larger-scale amplitude--angle state assigns greater probability to the
observed smaller-scale state than does a model containing only the
destination-state frequencies.

For horizons $h=1,\ldots,6$, the mean information gains are,
respectively, $1.820$, $0.983$, $0.602$, $0.401$, $0.284$, and
$0.212$ bits per transition. The predictive information decreases as
the scale separation increases but remains positive throughout the
tested range. The scale-dependent conditional structure measured in
Cube A therefore transfers to a spatially disjoint part of the same
flow without refitting. Because the two cubes belong to the same DNS
realization, this test establishes spatial transfer within that
realization rather than universality across independent turbulent
systems.

\section{Eulerian temporal measurements and nonlinear attribution}
\label{app:temporal_nonlinear}

The temporal analysis uses two distinct Eulerian constructions. For the
angular-persistence measurement, the local-perpendicular increment
geometry is reconstructed independently at each stored time. For the
nonlinear-term attribution, the endpoints and perpendicular plane
defined by the source-time geometry are held fixed when the destination
field is sampled. Neither construction advects the endpoints or follows
a material structure.

\subsection{Temporal sampling, angular persistence, and transfer}
\label{app:temporal_sampling_transfer}

The calculation uses sixty-five stored velocity and magnetic fields with
indices
$1,17,33,\ldots,1025$. The numerical time associated with stored index
$n$ is $t=0.0025n$, so the sixty-four consecutive intervals have
duration $\Delta t=0.04$. The same $4000$ spatial midpoints and eight
local-perpendicular direction labels are used at every time, giving
$32000$ increment pairs at each separation.

The temporal angular-persistence measurements use
$r=64$, $80$, $96$, and $112$. At every stored time, the locally
averaged magnetic field, perpendicular basis, physical sampling
direction, and centered increment endpoints are recalculated. A sample
is matched between two times only through its spatial midpoint and
direction label.

For two consecutive stored times $t_0=t$ and
$t_1=t+\Delta t$, let $\widehat{\mathbf b}_k$ denote the local
mean-field direction and define
\begin{align}
\mathbf P_k
&=
\mathbf I-
\widehat{\mathbf b}_k
\widehat{\mathbf b}_k^{\,T},
\nonumber\\
\mathbf q_k^\pm
&=
\mathbf P_k
\delta_{\mathbf r}\mathbf z^\pm(t_k),
\qquad
k=0,1 .
\label{eq:app_temporal_projected_states}
\end{align}
Here the centered increment at each $t_k$ is constructed using the
local-perpendicular geometry determined independently at that time.
The corresponding $A_k$, $s_k$, and $\theta_k$ are calculated from
$\mathbf q_k^\pm$ using the definitions in the main text.

Because the local geometry is reconstructed independently at the two
times, a measured change of mutual angle can contain a contribution
from the change of the sampling geometry. We quantify these changes by
\begin{align}
\alpha_r =
\arccos\!\left[
\left|
\widehat{\mathbf r}_j(t_0)\cdot
\widehat{\mathbf r}_j(t_1)
\right|
\right],
\qquad
\alpha_b =
\arccos\!\left[
\left|
\widehat{\mathbf b}_0\cdot
\widehat{\mathbf b}_1
\right|
\right].
\label{eq:app_temporal_geometry_rotations}
\end{align}
Here $\alpha_r$ is the principal angle between the two centered
sampling axes and $\alpha_b$ is the principal angle between the two
local perpendicular planes.  Absolute scalar products are used
because a centered sampling axis is unchanged by
$\widehat{\mathbf r}\rightarrow-\widehat{\mathbf r}$ and a
perpendicular plane is unchanged by
$\widehat{\mathbf b}\rightarrow-\widehat{\mathbf b}$; hence both
geometrical changes lie in $[0,\pi/2]$. We control for both quantities when measuring the
amplitude dependence of the subsequent mutual-angle change, using the
partial rank correlations
\begin{align}
\rho_s
&=
\rho_{\rm S}
\left[
\ln A_0,
|s_1-s_0|
\;\middle|\;
s_0,\alpha_r,\alpha_b
\right],
\nonumber\\
\rho_\theta
&=
\rho_{\rm S}
\left[
\ln A_0,
|\theta_1-\theta_0|
\;\middle|\;
\theta_0,\alpha_r,\alpha_b
\right].
\label{eq:app_temporal_partial_correlations}
\end{align}
All variables are first replaced by their ranks. Each of the two
variables whose association is being measured is then linearly
regressed on the ranked conditioning variables, and the reported
coefficient is the Pearson correlation between the two residuals.
Thus the procedure removes linear dependence in rank space without
assuming a parametric relation among the original variables.

The median interval-by-interval values are
\[
\begin{array}{c|cc}
r & \rho_s & \rho_\theta \\
\hline
64  & -0.3461 & -0.3572 \\
80  & -0.3398 & -0.3522 \\
96  & -0.3272 & -0.3425 \\
112 & -0.3183 & -0.3343
\end{array}
\]
and both correlations are negative in every one of the sixty-four
stored-time intervals at every separation, as summarized in
Table~\ref{tab:temporal_angular_persistence}. Thus larger initial
amplitude is associated with smaller subsequent angular change after
the initial angular state and changes of the local sampling geometry
have been accounted for. The agreement between the $s$-based and
direct-angle measurements also shows that this result is not produced
by the transformation $s=\sin\theta$.

As a stricter check that this result is genuinely conditional on
initial angle rather than a consequence of the global partial-rank
control, each source population was also divided into six
equal-population initial-angle ranges.  The partial-rank calculation
was then repeated separately inside every angular range, retaining
the residual control for the exact initial angle and for
$\alpha_r$ and $\alpha_b$.  There are therefore
$4\times64\times6=1536$ conditional tests for each angular measure.
The amplitude--angular-displacement correlation is negative in all
1536/1536 tests for $|s_1-s_0|$ and independently in all 1536/1536
tests for $|\theta_1-\theta_0|$.  The negative ordering holds in all
384/384 tests at each of the four separations and in all 256/256
tests within each of the six initial-angle ranges.  Thus the temporal
persistence result is unchanged when the initial angular population
is explicitly restricted before the amplitude dependence is
measured.

The same test was repeated in the spatially disjoint Cube B defined in
Appendix~\ref{app:sampling_ck}. The same sixty-five stored fields,
four separations, and numbers of midpoint and local-perpendicular
direction labels were used, with the complete local geometry again
reconstructed independently at every stored time. Both $\rho_s$ and
$\rho_\theta$ are negative in all sixty-four consecutive intervals at
each of the four separations. The temporal amplitude dependence of the
angular change is therefore not restricted to Cube A.

We separately test whether the full conditional amplitude--angle
dependence transfers between the two spatial regions. At each
separation the state
\[
\boldsymbol{\xi}_r(t)
=
\bigl(\ln A_r(t),s_r(t)\bigr)
\]
is divided into six equal-population ranges in each coordinate, giving
thirty-six joint states. The state boundaries are estimated from the
Cube-A training data and are then applied to Cube B without refitting.
For a stored-time lag $m\Delta t$, a separate direct finite-time kernel
is estimated,
\begin{align}
K^{A}_{\alpha\beta}(m;r)
&=
P_A
\left[
\boldsymbol{\xi}_r(t+m\Delta t)\in\alpha
\,\middle|\,
\boldsymbol{\xi}_r(t)\in\beta
\right].
\label{eq:app_temporal_transfer_kernel}
\end{align}
We use $m=1,2,4,8,$ and $16$. These are independently measured
finite-time kernels, not compositions of a one-step kernel; no temporal
Markov assumption enters this test.

The Cube-A kernel is evaluated on the measured Cube-B
source--destination pairs without changing its state boundaries or
transition probabilities. Its predictive log score is compared with
that of the unconditional Cube-A destination distribution, which
retains the later-state frequencies but discards the initial
amplitude--angle state. The resulting information gain is
\begin{align}
G(m)
&=
\frac{1}{N\ln 2}
\sum_{\ell=1}^{N}
\ln
\frac{
K^{A}_{\alpha_\ell\beta_\ell}(m;r)
}{
p^{A}_{{\rm dest},\alpha_\ell}(m;r)
}.
\label{eq:app_temporal_information_gain}
\end{align}
A positive value means that knowledge of the initial joint state
assigns greater probability to the later state actually measured in
Cube B than does the unconditional destination model.

For the strict time-holdout test, both the state boundaries and the
kernels are estimated using only the earlier Cube-A measurements and
are evaluated on later Cube-B source--destination pairs excluded from
the estimation. Averaged over the four separations and the available
held-out intervals, the information gains for
$m=1,2,4,8,$ and $16$ are
\[
G
=
1.893,\quad
1.047,\quad
0.502,\quad
0.237,\quad
0.106
\]
bits per transition. The transferred conditional information decreases
with lag but remains positive through the largest tested interval,
$16\Delta t=0.64$.

At short lags the Eulerian state itself is strongly persistent, so this
result should not be interpreted as showing that the transferred
kernel systematically outperforms a persistence model in
probability-distribution distance. It establishes instead that the
initial amplitude--angle state contains transferable predictive
information relative to a model that retains only the unconditional
later-state distribution. Since Cubes A and B belong to the same DNS
realization, the result establishes spatial and temporal transfer
within that realization rather than universality across independent
turbulent realizations.

\subsection{Fixed-endpoint nonlinear-term attribution}
\label{app:fixed_endpoint_attribution}

The nonlinear-term attribution uses the same sixty-four stored-time
intervals and is repeated at fourteen separations,
\[
r=
32,40,48,56,64,72,80,88,96,104,112,120,128,136 .
\]
For each source time and separation, the local-perpendicular sampling
direction defines the fixed endpoints
\begin{align}
\mathbf x_\pm
&=
\mathbf x
\pm
\frac{r}{2}
\widehat{\mathbf r}_j(t).
\label{eq:app_fixed_endpoints}
\end{align}
Let $\widehat{\mathbf b}_0$ be the local mean-field direction at the
source time and
\begin{align}
\mathbf P_0
&=
\mathbf I-
\widehat{\mathbf b}_0
\widehat{\mathbf b}_0^{\,T}.
\label{eq:app_source_projector}
\end{align}
The source increment and the measured destination increment are both
evaluated at the same endpoints and in this same perpendicular plane:
\begin{align}
\mathbf q^\pm_0
&=
\mathbf P_0
\left[
\mathbf z^\pm(\mathbf x_+,t)
-
\mathbf z^\pm(\mathbf x_-,t)
\right],
\nonumber\\
\mathbf q^\pm_{1,{\rm fix}}
&=
\mathbf P_0
\left[
\mathbf z^\pm(\mathbf x_+,t+\Delta t)
-
\mathbf z^\pm(\mathbf x_-,t+\Delta t)
\right].
\label{eq:app_fixed_endpoint_states}
\end{align}
This construction removes changes caused solely by rotation of the
local sampling plane or displacement of the increment endpoints.

The local Els\"asser advective term and its increment between the
fixed endpoints are
\begin{align}
\mathbf N^\pm
&=
-
\left(
\mathbf z^\mp\mathbin{\cdot}\boldsymbol{\nabla}
\right)
\mathbf z^\pm,
\nonumber\\
\mathbf R^\pm_{\rm NL}(t)
&=
\mathbf N^\pm(\mathbf x_+,t)
-
\mathbf N^\pm(\mathbf x_-,t).
\label{eq:app_nonlinear_increment}
\end{align}
Its finite-time contribution is estimated by the trapezoidal vector
update
\begin{align}
\mathbf q^\pm_{1,{\rm NL}}
&=
\mathbf q^\pm_0
+
\frac{\Delta t}{2}
\mathbf P_0
\left[
\mathbf R^\pm_{\rm NL}(t)
+
\mathbf R^\pm_{\rm NL}(t+\Delta t)
\right].
\label{eq:app_trapezoidal_update}
\end{align}
The same projector $\mathbf P_0$ is therefore applied to the source
increment, the measured destination increment, and the nonlinear term
estimate. The destination values of $A$, $a=\ln A$, and $s$ are
calculated directly from the resulting vectors rather than from a
linearized scalar update.

The velocity and magnetic fields are evaluated at the generally
off-grid endpoints by trilinear interpolation. Spatial derivatives are
first calculated at grid points with fourth-order centered differences
and are then interpolated to the same endpoints. The source field,
measured destination field, and nonlinear term estimate therefore use
the same endpoint coordinates and interpolation convention.

For every interval and separation, the resolved source
$(a_0,s_0)$ plane is partitioned adaptively into approximately
equal-population cells. The source values of $a_0$ are first divided
into $n$ equal-population ranges; within each amplitude range, $s_0$ is
divided into another $n$ equal-population ranges. We use
$n=6$, $8$, and $10$. The resulting source-time boundaries are then
held fixed when the measured fixed-endpoint and nonlinear-estimated
destination states are classified.

For source cell $\beta$ and destination cell $\alpha$, define
\begin{align}
K^{Y}_{\alpha\beta}
&=
P
\left(
\boldsymbol{\xi}_{1,Y}\in\alpha
\,\middle|\,
\boldsymbol{\xi}_0\in\beta
\right),
\qquad
Y\in\{{\rm fix},{\rm NL}\},
\nonumber\\
D_{\rm TV}^{\rm NL}
&=
\frac{1}{2}
\sum_\beta p_\beta
\sum_\alpha
\left|
K^{\rm fix}_{\alpha\beta}
-
K^{\rm NL}_{\alpha\beta}
\right|,
\label{eq:app_nonlinear_kernel_distance}
\end{align}
where $p_\beta$ is the measured source-cell probability. The latter is
the transition-kernel distance used in
Fig.~\ref{fig:multiscale_nonlinear_attribution}.

The corresponding conditional mean changes are
\begin{align}
\mu_a^Y(\beta)
&=
\left\langle
a_{1,Y}-a_0
\mid\beta
\right\rangle,
&
\mu_s^Y(\beta)
&=
\left\langle
s_{1,Y}-s_0
\mid\beta
\right\rangle .
\label{eq:app_nonlinear_conditional_fields}
\end{align}
We also evaluate the conditional fields of
$|\Delta s|$ and $(\Delta s)^2$. Comparisons between measured and
nonlinear-estimated conditional fields are weighted by the number of
source measurements in each cell.

The permutation control randomly reassigns the nonlinear-estimated
destination pairs
$(a_{1,{\rm NL}},s_{1,{\rm NL}})$ among the source labels within each
stored-time interval, while leaving the measured source states and
source-cell boundaries unchanged. It therefore preserves the complete
one-time distribution of the nonlinear estimate but destroys its
association with the source state. Twenty independent permutations are
used for every interval and state-space partition.

For the primary $10\times10$ partition, after averaging over the
sixty-four intervals at each separation, the transition-kernel distance
between the measured evolution and the nonlinear term estimate lies
between $0.122$ and $0.150$ over the fourteen tested separations. The
corresponding permutation control lies between $0.764$ and $0.817$.
There is no systematic deterioration of the nonlinear term estimate
toward either end of the tested separation range.

The conditional-field agreement is still closer. Across the same
fourteen separations, the conditional-field correlation between the
measured and nonlinear-estimated $\Delta\ln A$ lies between $0.923$
and $0.977$, while that for $\Delta s$ lies between $0.988$ and
$0.994$. These are the two conditional-field correlations shown in
Fig.~\ref{fig:multiscale_nonlinear_attribution}. The result therefore does not depend on a single representative
separation: the local Els\"asser advective term shows strong and
reproducible agreement with the resolved state dependence of both
amplitude and angular change throughout the tested range.

As a secondary control, the source population is also separated into
large-amplitude/small-angle (LS) and large-amplitude/large-angle (LL)
states. For an upper-amplitude fraction $p$, the source thresholds are
\begin{align}
A_p
&=
Q_{1-p}[A_0],
\nonumber\\
s_L
&=
Q_{1/3}
\left[
s_0\mid A_0\geq A_p
\right],
\qquad
s_U
=
Q_{2/3}
\left[
s_0\mid A_0\geq A_p
\right],
\nonumber\\
{\rm LS}
&:
A_0\geq A_p,\quad s_0\leq s_L,
\qquad
{\rm LL}
:
A_0\geq A_p,\quad s_0\geq s_U .
\label{eq:app_hs_hl_source_classes}
\end{align}
The calculation is repeated for $p=0.05$, $0.10$, and $0.20$.
Before departure probabilities are compared, the source
distributions of $a_0=\ln A_0$ in the two classes are reweighted to a
common distribution over their shared support. If
$P_{{\rm LS},b}$ and $P_{{\rm LL},b}$ are their normalized masses in
amplitude interval $b$, the common target distribution is proportional
to
\begin{align}
m_b
&=
\min
\left(
P_{{\rm LS},b},
P_{{\rm LL},b}
\right).
\label{eq:app_hs_hl_amplitude_match}
\end{align}
The primary comparison uses twenty amplitude intervals, with nearby
choices used as robustness checks.

After this source-amplitude matching, the difference between the LL
and LS departure probabilities does not retain a universal sign
through the temporal record. The categorical result is therefore not
used as the primary nonlinear-attribution statistic. Instead, it shows that the two-class departure behavior does not support a universal angle-only departure law at fixed source amplitude. A large-amplitude large-angle state can,
for example, leave the large-amplitude population through amplitude
loss while retaining a large mutual angle.

Finally, Eq.~(\ref{eq:app_trapezoidal_update}) is a finite-time
term-attribution construction, not an autonomous nonlinear prediction.
In particular,
$\mathbf R^\pm_{\rm NL}(t+\Delta t)$ is evaluated from the actual
destination field produced by the complete MHD evolution. The
difference between
$\mathbf q^\pm_{1,{\rm fix}}$ and
$\mathbf q^\pm_{1,{\rm NL}}$ therefore contains the effects of
pressure, dissipation, forcing, finite-time quadrature, spatial
differentiation, interpolation, and other omitted contributions. The calculation establishes that the measured local Els\"asser
advective term closely tracks the conditional state dependence of
the finite-time amplitude and angular evolution; it does not imply
that this term alone generates the complete temporal change.

\section{Reduced-MHD numerical details and robustness tests}
\label{app:beresnyak}

The independent scale-space test uses the balanced strong-guide-field
reduced-MHD simulations described by
\citet{Beresnyak2014,Beresnyak2015}. We use the $1024^3$
ordinary-viscosity run \texttt{b1024n}. The original angular-persistence analysis is based on the stored state
\texttt{R108000}, with \texttt{R162000} used as an additional
reproducibility test. The higher-order one-scale statistics and the
Politano--Pouquet calculation use the five stored states
\texttt{R018000}, \texttt{R072000}, \texttt{R108000},
\texttt{R126000}, and \texttt{R162000}.

The available fields are the four transverse components
$v_y$, $v_z$, $b_y$, and $b_z$, from which the perpendicular
Els\"asser fields are constructed as
$\boldsymbol z_\perp^\pm=\boldsymbol v_\perp\pm\boldsymbol b_\perp$.
Unlike the full-MHD calculation, the perpendicular plane does not have
to be reconstructed from a local magnetic field: the RMHD fluctuations
are already transverse to the imposed guide field. Taking the guide
field along the $x$ direction, all increments are therefore measured
in the fixed $y$--$z$ plane.

The primary stored state is close to balanced. Direct volume averages
give normalized cross helicity $\sigma_c=0.0171$, normalized residual
energy $\sigma_r=-0.1219$, and $E^+/E^-=1.0348$. Thus the two
Els\"asser populations have nearly equal energy, while the negative
residual energy indicates a modest excess of magnetic over kinetic
fluctuation energy.

Centered perpendicular increments are evaluated with periodic boundary
conditions, using the same definitions of $A_r$, $s_r$, and
$\theta_r$ as in the main text. When two separations are compared, the
same spatial midpoint and the same perpendicular direction are retained.
The sampling uses two interlaced regular sets of midpoints with grid
spacing $16$. This gives $524288$ matched measurements for each of the
two orthogonal perpendicular directions and $1048576$ measurements
when the two directions are pooled.

We use nine scale pairs,
\[
r_>=32,48,64,80,96,112,128,144,160,
\qquad
r_<=\frac{7}{8}r_>.
\]
All centered endpoints lie on grid points, so no spatial interpolation
is required. At each larger separation, the measurements are divided
into six equal-population amplitude ranges in $a_>=\ln A_{r_>}$ and,
independently, six equal-population ranges in $s_>$. The angular
changes are then measured with $M_s^{(2)}$ and
$M_\theta^{(2)}$ at fixed initial angular range. This conditioning is
important: the test asks whether measurements beginning with comparable
angles undergo different angular changes according to their initial
amplitude.

For the primary state \texttt{R108000}, the amplitude ordering is
uniform over the tested scale range. At fixed initial angular range,
$M_s^{(2)}$ decreases through every successive amplitude range in all
$270/270$ adjacent-amplitude comparisons for the $y$-directed
separations and independently in all $270/270$ comparisons for the
$z$-directed separations. The direct folded-angle measure
$M_\theta^{(2)}$ gives the same $270/270$ ordering in each direction.
Pooling the two perpendicular directions gives the same result. The
amplitude contrast becomes weaker toward the largest separations but
remains present throughout the tested range.

We also tested whether this ordering is produced by a restricted part
of the RMHD volume. The $1024^3$ domain was divided, according to
midpoint position, into eight disjoint $512^3$ octants. The amplitude
and angular boundaries obtained from the complete stored state were
kept fixed when the octants were analyzed, so each spatial subset was
tested against the same source-state definition. Three representative
transitions,
$32\rightarrow28$, $96\rightarrow84$, and
$160\rightarrow140$, were evaluated separately for both perpendicular
directions.
\begin{table}
\centering
\caption{
Summary of the RMHD scale-space angular-persistence tests.
``Adjacent'' counts comparisons between successive amplitude ranges at
fixed initial angular range. ``Large$<$Small'' counts cases in which the
largest-amplitude range has smaller angular displacement than the
smallest-amplitude range.
}
\label{tab:beresnyak_robustness}
\footnotesize
\setlength{\tabcolsep}{4pt}

\resizebox{0.97\textwidth}{!}{%
\begin{tabular}{@{}lccc@{}}
\hline
Test & Coverage & $M_s^{(2)}$ & $M_\theta^{(2)}$ \\
\hline
\texttt{R108000}, $y$ direction
& 9 scale pairs
& 270/270 adjacent
& 270/270 adjacent \\

\texttt{R108000}, $z$ direction
& 9 scale pairs
& 270/270 adjacent
& 270/270 adjacent \\

\texttt{R108000}, spatial octants
& 3 pairs, 8 octants, 2 directions
& 288/288 large$<$small; 1429/1440 adjacent
& 288/288 large$<$small; 1431/1440 adjacent \\

\texttt{R162000}
& 3 pairs, 2 directions
& 36/36 large$<$small; 180/180 adjacent
& 36/36 large$<$small; 180/180 adjacent \\
\hline
\end{tabular}%
}
\end{table}
In every one of the resulting $288$ scale--direction--angle
comparisons, the largest-amplitude population has smaller
$M_s^{(2)}$ than the smallest-amplitude population. The same
$288/288$ ordering is obtained for $M_\theta^{(2)}$. The stronger
requirement that the angular displacement decrease between every pair
of successive amplitude ranges is satisfied in $1429/1440$
comparisons for $M_s^{(2)}$ and $1431/1440$ comparisons for
$M_\theta^{(2)}$. The persistence ordering is therefore distributed
throughout the RMHD volume rather than being generated by one spatial
region.

The same three representative transitions were then repeated using the
separate stored state \texttt{R162000}. The two perpendicular
directions were again treated independently. For both angular measures,
the largest-amplitude population has smaller angular displacement than
the smallest-amplitude population in all $36/36$
scale--direction--angle comparisons. Moreover, the angular displacement
decreases through every successive amplitude range in all $180/180$
adjacent-amplitude comparisons for both $M_s^{(2)}$ and
$M_\theta^{(2)}$.

The higher-order one-scale statistics shown in
Fig.~\ref{fig:higher_moment_alignment} are evaluated independently in
the five stored states \texttt{R018000}, \texttt{R072000},
\texttt{R108000}, \texttt{R126000}, and \texttt{R162000}. We use
\[
r=32,\ 48,\ 64,\ 80,\ 96,\ 112,\ 128,\ 144,\ 160.
\]
For this calculation we use a single regular midpoint lattice with grid
spacing $16$, distinct from the two-interlaced-lattice sampling used
for the angular-persistence test above. This gives $64^3=262144$
midpoints for each perpendicular increment direction and $524288$
measurements after the two directions are pooled, per stored state and
separation. For each stored
state we calculate $I_2(r)$, $I_3(r)$, and
$\langle s_r\rangle_{A^p}$ for $p=0,1,2,3$. The plotted values are
the means across the five stored states, with standard errors estimated
across those states rather than across the individual spatial samples.

Because the $p=3$ quantities give substantial weight to rare
large-amplitude fluctuations, we also repeated the calculation with a
denser spatial lattice in \texttt{R108000} and \texttt{R162000} at
$r=32$, $64$, and $128$. Reducing the grid spacing from $16$ to $8$
increases the pooled sample to $4194304$ measurements per stored state
and separation. Relative changes in $I_3$ range from $0.21\%$ to
$2.16\%$, while the absolute change in
$\langle s_r\rangle_{A^3}$ is at most $1.0\times10^{-3}$.
The ordering
\[
\langle s_r\rangle_{A^0}>
\langle s_r\rangle_A>
\langle s_r\rangle_{A^2}>
\langle s_r\rangle_{A^3}
\]
is unchanged in all six dense-sampling comparisons.

The exact finite-step decomposition in Eq.~(\ref{eq:weighted_finite_step_decomposition}) gives a
stronger version of this comparison.  For all nine RMHD scale
pairs, the amplitude-weighted statistic decreases toward the
smaller separation.  In all 9/9 transitions the joint term
$\langle\Delta A\,\Delta s\rangle/\langle A_<\rangle$ is
negative and is the largest of the three terms in absolute
magnitude, whereas neither the direct angular term nor the
pure amplitude-reweighting term is negative.  The median
fractions of the sum of the absolute magnitudes of the direct,
reweighting, and joint terms are $0.20$, $0.24$, and $0.56$,
respectively.

The same ordering is recovered when the two perpendicular
directions and eight spatial octants are treated separately.
For the direct $160\rightarrow28$ comparison, the joint term
is the largest contribution in 13/16 direction--octant
subsamples.  Its interquartile range is
$[-0.191,-0.116]$, whereas that of the pure
amplitude-reweighting term is positive,
$[0.029,0.053]$.  The RMHD calculation therefore independently shows that the decrease of
the weighted alignment statistic requires the joint amplitude--angle
change during the same scale transition.

\subsection{Politano--Pouquet calculation}
\label{app:pp_calculation}

The Politano--Pouquet calculation is performed separately in the five
stored states \texttt{R018000}, \texttt{R072000}, \texttt{R108000},
\texttt{R126000}, and \texttt{R162000}. Centered increments are
evaluated at $2{,}097{,}152$ midpoint locations for each perpendicular
separation direction and each separation. These locations are spatial
samples and are not treated as statistically independent realizations.
The $y$- and $z$-directed measurements and the $+$ and $-$ Els\"asser
relations are first analyzed separately.

The unconditioned third-order moment is measured at perpendicular
separations
\[
28,32,42,48,56,64,70,80,84,96,98,112,126,128,140,144,160
\]
grid points. Its sign and proportionality to separation are tested
through the scale dependence of
Eq.~(\ref{eq:pp_scale_estimate}). Grid-point separations are sufficient
for testing this scale dependence; no absolute value of
$\varepsilon^\pm$ is inferred from their numerical units.

The conditional calculation uses the nine source scales
\[
32,48,64,80,96,112,128,144,160.
\]
Angular changes are evaluated for all $36$ ordered pairs of these scales
satisfying $r_<<r_>$. The source state is partitioned using four, six,
or eight equal-population angular ranges. Within each angular range,
the amplitude is divided into the same number of equal-population
ranges. The comparison of angular displacement between amplitude
populations is therefore always made at fixed source-angle range. The
population-weighted conditional PP means are summed and compared with
the directly measured third-order moment. Separate perpendicular
directions, spatial octants, state-space resolutions, and stored states
provide the numerical robustness tests.

Across the five stored states and the two Els\"asser relations, the
third-order moment has the negative sign required for transfer toward
smaller perpendicular scales in $2654/2720$ separate
separation--direction--octant estimates. The coefficient of variation
with separation of the normalized PP quantity ranges from $0.0402$ to
$0.2511$ among the ten stored-state--Els\"asser combinations. The
median relative difference between the $y$- and $z$-directed
measurements ranges from $0.093$ to $0.836$. Thus the sign is highly
reproducible, while the magnitude of the third-order moment is less
well converged in some stored states and directions.

The conditional-state sums reproduce the directly measured PP moments
with a largest relative difference of $1.13\times10^{-15}$. For the
largest-angle, highest-amplitude source population, the mean of the
$+$ and $-$ conditional PP moments is negative at all nine source
scales in every stored state. This gives $135/135$ tests across the
five stored states and the $4\times4$, $6\times6$, and $8\times8$
state-space partitions.

The corresponding angular-persistence ordering is also insensitive to
the state-space resolution. The highest-amplitude range has smaller
$M_\theta^{(2)}$ than the lowest-amplitude range in $720/720$
scale-pair--angle comparisons for the $4\times4$ partitions,
$1074/1080$ comparisons for the $6\times6$ partitions, and
$1424/1440$ comparisons for the $8\times8$ partitions. Combining the
three resolutions gives $3218/3240$, or $99.3\%$, of all comparisons.

These tests show that the scale-space result is not
specific to a single separation, perpendicular direction, spatial
region, or stored state. At fixed initial angle, large-amplitude
Els\"asser-increment pairs undergo smaller angular changes across scale,
including when they begin at large angles. The independent RMHD calculation therefore shows the same amplitude
ordering. A fixed vector change turns a larger increment through a smaller
angle, so this ordering can arise from geometry and does not by itself
establish a universal one-way rotation toward alignment.

\section{Carr\'e du champ form of the infinitesimal cross term}
\label{app:carre_du_champ}

This appendix records the infinitesimal counterpart of the finite-step
decomposition in
Eq.~(\ref{eq:weighted_finite_step_decomposition}). No continuous
generator is assumed in the numerical results of this paper.

Suppose that the reduced amplitude--angle state admits a continuous
Markov evolution in $\tau$, with generator ${\cal L}_\tau$ acting on
observables. For any sufficiently regular observable $F$,
\begin{align}
\frac{d}{d\tau}\langle F\rangle
=
\left\langle{\cal L}_\tau F\right\rangle .
\label{eq:carre_observable}
\end{align}
The carr\'e du champ operator associated with ${\cal L}_\tau$ is
defined by \cite{BakryGentilLedoux2014}
\begin{align}
\Gamma_\tau(F,G)
=
\frac{1}{2}
\left[
{\cal L}_\tau(FG)
-
F{\cal L}_\tau G
-
G{\cal L}_\tau F
\right].
\label{eq:carre_definition}
\end{align}
It measures the failure of the generator to obey the ordinary product
rule. Applying Eq.~(\ref{eq:carre_definition}) to $A=e^a$ and $s$
gives
\begin{align}
{\cal L}_\tau(As)
=
A{\cal L}_\tau s
+
s{\cal L}_\tau A
+
2\Gamma_\tau(A,s).
\label{eq:carre_product}
\end{align}
Using Eq.~(\ref{eq:carre_observable}) and differentiating
$\langle s\rangle_A=\langle As\rangle/\langle A\rangle$ therefore
gives
\begin{align}
\frac{d}{d\tau}\langle s\rangle_A
=
\frac{1}{\langle A\rangle}
\bigg[
&
\left\langle A{\cal L}_\tau s\right\rangle
+
\left\langle
\bigl(s-\langle s\rangle_A\bigr){\cal L}_\tau A
\right\rangle
\nonumber\\
&
+
2\left\langle\Gamma_\tau(A,s)\right\rangle
\bigg].
\label{eq:carre_weighted_alignment}
\end{align}
The three terms are the infinitesimal counterparts of the angular,
amplitude-reweighting, and cross terms in
Eq.~(\ref{eq:weighted_finite_step_decomposition}). In particular,
under the usual conditions for the short-step limit,
\begin{align}
\Gamma_\tau(A,s)
=
\lim_{\Delta\tau\rightarrow0}
\frac{
\left\langle
\Delta A\,\Delta s
\mid A,s
\right\rangle
}{
2\Delta\tau
}.
\label{eq:carre_short_step}
\end{align}
Thus $\Gamma_\tau(A,s)$ measures the joint infinitesimal variation of
amplitude and angle. It vanishes for a deterministic first-order
generator, which obeys the ordinary product rule, but can be nonzero
for diffusion or jump processes. The present analysis does not
estimate the limit in Eq.~(\ref{eq:carre_short_step}); its numerical
conclusions rely on the finite-step identity
Eq.~(\ref{eq:weighted_finite_step_decomposition}).

\bibliography{AngularPersistence}

@book{BakryGentilLedoux2014,
  author    = {Dominique Bakry and Ivan Gentil and Michel Ledoux},
  title     = {Analysis and Geometry of Markov Diffusion Operators},
  series    = {Grundlehren der mathematischen Wissenschaften},
  volume    = {348},
  publisher = {Springer},
  address   = {Cham},
  year      = {2014},
  doi       = {10.1007/978-3-319-00227-9}
}

@article{PolitanoPouquet1998,
  author  = {Politano, H. and Pouquet, A.},
  title   = {von K{\'a}rm{\'a}n--Howarth equation for
             magnetohydrodynamics and its consequences on third-order
             longitudinal structure and correlation functions},
  journal = {Physical Review E},
  volume  = {57},
  pages   = {R21--R24},
  year    = {1998},
  doi     = {10.1103/PhysRevE.57.R21}
}

@article{Galtier2011,
  author  = {Galtier, S.},
  title   = {Third-order Els{\"a}sser moments in axisymmetric
             MHD turbulence},
  journal = {Comptes Rendus Physique},
  volume  = {12},
  pages   = {151--159},
  year    = {2011},
  doi     = {10.1016/j.crhy.2010.11.006}
}

@article{Beresnyak2012,
  author  = {Beresnyak, Andrey},
  title   = {Basic properties of magnetohydrodynamic turbulence in the inertial range},
  journal = {Monthly Notices of the Royal Astronomical Society},
  year    = {2012},
  volume  = {422},
  number  = {4},
  pages   = {3495--3502},
  doi     = {10.1111/j.1365-2966.2012.20859.x},
  eprint  = {1111.5329},
  archivePrefix = {arXiv}
}

@article{Beresnyak2015,
  author        = {Beresnyak, Andrey},
  title         = {On the Parallel Spectrum in Magnetohydrodynamic Turbulence},
  journal       = {The Astrophysical Journal Letters},
  volume        = {801},
  number        = {1},
  pages         = {L9},
  year          = {2015},
  doi           = {10.1088/2041-8205/801/1/L9},
  eprint        = {1407.2613},
  archivePrefix = {arXiv},
  primaryClass  = {astro-ph.SR}
}

@article{Beresnyak2014,
  author  = {Beresnyak, Andrey},
  title   = {Spectra of Strong Magnetohydrodynamic Turbulence from High-resolution Simulations},
  journal = {The Astrophysical Journal Letters},
  volume  = {784},
  number  = {2},
  pages   = {L20},
  year    = {2014},
  doi     = {10.1088/2041-8205/784/2/L20},
  eprint  = {1401.4177},
  archivePrefix = {arXiv},
  primaryClass  = {astro-ph.GA}
}

@article{MalletSchekochihin2017,
  author  = {Mallet, A. and Schekochihin, A. A.},
  title   = {A Statistical Model of Three-dimensional Anisotropy and
             Intermittency in Strong Alfv{\'e}nic Turbulence},
  journal = {Monthly Notices of the Royal Astronomical Society},
  year    = {2017},
  volume  = {466},
  number  = {4},
  pages   = {3918--3927},
  doi     = {10.1093/mnras/stw3251}
}

@article{MalletSchekochihinChandran2017,
  author  = {Mallet, A. and Schekochihin, A. A. and Chandran, B. D. G.},
  title   = {Disruption of Sheet-like Structures in Alfv{\'e}nic
             Turbulence by Magnetic Reconnection},
  journal = {Monthly Notices of the Royal Astronomical Society},
  year    = {2017},
  volume  = {468},
  number  = {4},
  pages   = {4862--4871},
  doi     = {10.1093/mnras/stx670}
}

@article{SioulasEtAl2025,
  author  = {Sioulas, Nikos and Zikopoulos, Themistocles and Shi, Chen
             and Velli, Marco and Bowen, Trevor A. and Mallet, Alfred
             and Chandran, Benjamin D. G. and Sorriso-Valvo, Luca
             and Martinovi{\'c}, Mihailo M. and Cerri, Silvio S.
             and Verdini, Andrea and Davis, Nooshin and Dunn, Corina},
  title   = {Higher-order Analysis of Three-dimensional Anisotropy in
             Imbalanced Alfv{\'e}nic Turbulence},
  journal = {The Astrophysical Journal},
  year    = {2025},
  volume  = {993},
  number  = {1},
  eid     = {142},
  pages   = {142},
  doi     = {10.3847/1538-4357/ae0934}
}

@article{BoldyrevMasonCattaneo2009,
  author  = {Boldyrev, Stanislav and Mason, Joanne and Cattaneo, Fausto},
  title   = {Dynamic Alignment and Exact Scaling Laws in Magnetohydrodynamic Turbulence},
  journal = {The Astrophysical Journal Letters},
  volume  = {699},
  number  = {1},
  pages   = {L39--L42},
  year    = {2009},
  doi     = {10.1088/0004-637X/699/1/L39}
}

@article{ChandranSchekochihinMallet2015,
  author  = {Chandran, Benjamin D. G. and Schekochihin, Alexander A. and Mallet, Alfred},
  title   = {Intermittency and Alignment in Strong RMHD Turbulence},
  journal = {The Astrophysical Journal},
  volume  = {807},
  number  = {1},
  pages   = {39},
  year    = {2015},
  doi     = {10.1088/0004-637X/807/1/39}
}

@article{MalletSchekochihinChandran2015,
  author  = {Mallet, A. and Schekochihin, A. A. and Chandran, B. D. G.},
  title   = {Refined Critical Balance in Strong Alfv{\'e}nic Turbulence},
  journal = {Monthly Notices of the Royal Astronomical Society: Letters},
  volume  = {449},
  number  = {1},
  pages   = {L77--L81},
  year    = {2015},
  doi     = {10.1093/mnrasl/slv021}
}

@article{MalletEtAl2016,
  author  = {Mallet, A. and Schekochihin, A. A. and Chandran, B. D. G. and Chen, C. H. K. and Horbury, T. S. and Wicks, R. T. and Greenan, C. C.},
  title   = {Measures of Three-Dimensional Anisotropy and Intermittency in Strong Alfv{\'e}nic Turbulence},
  journal = {Monthly Notices of the Royal Astronomical Society},
  volume  = {459},
  number  = {2},
  pages   = {2130--2139},
  year    = {2016},
  doi     = {10.1093/mnras/stw802}
}

@misc{SioulasEtAl2024,
  author        = {Sioulas, Nikos and Velli, Marco and Mallet, Alfred and Bowen, Trevor A. and Chandran, B. D. G. and Shi, Chen and Cerri, S. S. and Liodis, Ioannis and Ervin, Tamar and Larson, Davin E.},
  title         = {Scale-Dependent Dynamic Alignment in MHD Turbulence: Insights into Intermittency, Compressibility, and Imbalance Effects},
  year          = {2024},
  eprint        = {2407.03649},
  archivePrefix = {arXiv},
  primaryClass  = {physics.space-ph},
  doi           = {10.48550/arXiv.2407.03649}
}

@article{Beresnyak2011,
  title = {Spectral Slope and Kolmogorov Constant of MHD Turbulence},
  author = {Beresnyak, A.},
  journal = {Phys. Rev. Lett.},
  volume = {106},
  issue = {7},
  pages = {075001},
  numpages = {4},
  year = {2011},
  month = {Feb},
  publisher = {American Physical Society},
  doi = {10.1103/PhysRevLett.106.075001},
  url = {https://link.aps.org/doi/10.1103/PhysRevLett.106.075001}
}

@article{Schekochihin2022,
  title   = {MHD Turbulence: A Biased Review},
  author  = {Schekochihin, Alexander A.},
  journal = {Journal of Plasma Physics},
  volume  = {88},
  number  = {5},
  pages   = {155880501},
  year    = {2022},
  doi     = {10.1017/S0022377822000721}
}

@article{Mason2006,
  title = {Dynamic Alignment in Driven Magnetohydrodynamic Turbulence},
  author = {Mason, Joanne and Cattaneo, Fausto and Boldyrev, Stanislav},
  journal = {Phys. Rev. Lett.},
  volume = {97},
  issue = {25},
  pages = {255002},
  numpages = {4},
  year = {2006},
  month = {Dec},
  publisher = {American Physical Society},
  doi = {10.1103/PhysRevLett.97.255002},
  url = {https://link.aps.org/doi/10.1103/PhysRevLett.97.255002}
}

@article{Schekochihin2009,
doi = {10.1088/0067-0049/182/1/310},
url = {https://doi.org/10.1088/0067-0049/182/1/310},
year = {2009},
month = {may},
publisher = {The American Astronomical Society},
volume = {182},
number = {1},
pages = {310},
author = {Schekochihin, A. A. and Cowley, S. C. and Dorland, W. and Hammett, G. W. and Howes, G. G. and Quataert, E. and Tatsuno, T.},
title = {ASTROPHYSICAL GYROKINETICS: KINETIC AND FLUID TURBULENT CASCADES IN MAGNETIZED WEAKLY COLLISIONAL PLASMAS},
journal = {The Astrophysical Journal Supplement Series}
}

@article{Boldyrev2006,
  title = {Spectrum of Magnetohydrodynamic Turbulence},
  author = {Boldyrev, Stanislav},
  journal = {Phys. Rev. Lett.},
  volume = {96},
  issue = {11},
  pages = {115002},
  numpages = {4},
  year = {2006},
  month = {Mar},
  publisher = {American Physical Society},
  doi = {10.1103/PhysRevLett.96.115002},
  url = {https://link.aps.org/doi/10.1103/PhysRevLett.96.115002}
}

@article{Eyinketal2013,
	adsurl = {https://ui.adsabs.harvard.edu/abs/2013Natur.497..466E},
	author = {{Eyink}, G. and {Vishniac}, E. and {Lalescu}, C. and {Aluie}, H. and {Kanov}, K. and {Bürger}, K. and {Burns}, R. and {Meneveau}, C. and {Szalay}, A.},
	doi = {10.1038/nature12128},
	journal = {Nature},
	month = {may},
	pages = {466–469},
	title = {{Flux-freezing breakdown in high-conductivity magnetohydrodynamic turbulence}},
	volume = {497},
	year = {2013}
}

@article{JHTB1,
	Adsurl = {https://ui.adsabs.harvard.edu/abs/2008JTurb...9...31L},
	Archiveprefix = {arXiv},
	Author = {{Li}, Y. and {Perlman}, E. and {Wan}, M. and {Yang}, Y. and {Meneveau}, C. and {Burns}, R. and {Chen}, S. and {Szalay}, A. and {Eyink}, G.},
	Doi = {10.1080/14685240802376389},
	Eid = {N31},
	Eprint = {0804.1703},
	Journal = {Journal of Turbulence},
	Pages = {N31},
	Primaryclass = {physics.flu-dyn},
	Title = {{A public turbulence database cluster and applications to study Lagrangian evolution of velocity increments in turbulence}},
	Volume = {9},
	Year = {2008}}

@inproceedings{JHTB2,
author = {Perlman, Eric and Burns, Randal and Li, Yi and Meneveau, Charles},
title = {Data exploration of turbulence simulations using a database cluster},
year = {2007},
isbn = {9781595937643},
publisher = {Association for Computing Machinery},
address = {New York, NY, USA},
url = {https://doi.org/10.1145/1362622.1362654},
doi = {10.1145/1362622.1362654},
booktitle = {Proceedings of the 2007 ACM/IEEE Conference on Supercomputing},
articleno = {23},
numpages = {11},
location = {Reno, Nevada},
doi       = {10.1145/1362622.1362654},
series = {SC '07}
}
\end{document}